\documentclass[10pt,conference]{IEEEtran}
\IEEEoverridecommandlockouts

\usepackage[utf8]{inputenc}
\usepackage{booktabs}
\usepackage{moreverb}
\usepackage{fontenc}
\usepackage{amsmath}
\usepackage{amssymb}
\usepackage{fancybox}
\usepackage{color}
\usepackage{colortbl}
\usepackage{array}
\usepackage{multirow}
\usepackage{multicol}
\usepackage{listings}
\usepackage{graphicx}
\usepackage{wrapfig}
\usepackage[multiple,symbol]{footmisc}

\usepackage{subfloat}
\usepackage[captionskip=1pt,farskip=1pt,position=below]{subfig}

\usepackage{boxedminipage}

\usepackage[ruled,vlined,lined,commentsnumbered]{algorithm2e}
\usepackage{balance}
\usepackage{csvsimple}
\usepackage{longtable}
\usepackage{booktabs}
\usepackage[skip=1pt,labelfont=bf]{caption}
\usepackage{url}
\usepackage{lscape}
\usepackage{rotating}
\usepackage{tikz}
\usepackage[normalem]{ulem}
\usepackage{enumitem}
\usepackage{setspace}
\usepackage{hanging}
\usepackage{pdfpages}

\usepackage[most]{tcolorbox}

\usepackage{threeparttable}

\usepackage{marvosym}
\usepackage{soul}

\usepackage{enumitem}

\setlist{noitemsep} 

\definecolor{codegreen}{rgb}{0,0.6,0}
\definecolor{codegray}{rgb}{0.5,0.5,0.5}
\definecolor{codepurple}{rgb}{0.58,0,0.82}
\definecolor{backcolour}{rgb}{0.95,0.95,0.92}
\definecolor{yellow}{RGB}{255,255,153}
\definecolor{grey}{RGB}{224,224,224}

\newboolean{showcomments}
\setboolean{showcomments}{true}
\ifthenelse{\boolean{showcomments}}
 { \newcommand{\mynote}[2]{
      \fbox{\bfseries\sffamily\scriptsize#1}
        {\small$\blacktriangleright$\textsf{\emph{#2}}$\blacktriangleleft$}}}
        { \newcommand{\mynote}[2]{}}

\definecolor{DarkOrange}{rgb}{0.8,0.3,0.0}

\newcommand{\fixpattern}[1]{
\vspace{-0.6cm}
\begin{tcolorbox}[tile,size=fbox,boxsep=1mm,boxrule=0pt,top=0pt,bottom=0pt,
borderline west={2mm}{0pt}{black!5!white},colback=black!5!white, width=\linewidth] 
\ttfamily #1
\end{tcolorbox}
}

\newcommand{\find}[1]{
\vspace{-0.2cm}
\begin{tcolorbox}[tile,size=fbox,boxsep=2mm,boxrule=0pt,top=0pt,bottom=0pt,
borderline west={1mm}{0pt}{blue!50!white},colback=blue!5!white]
\em #1
\end{tcolorbox}
\vspace{-0.2cm}
}

\newcommand{\myquote}[1]{
\begin{tcolorbox}[tile,size=fbox,boxsep=0.4mm,boxrule=0pt,top=0pt,bottom=0pt,
borderline west={0.5mm}{0pt}{blue!5!white},colback=blue!5!white]
\footnotesize\em #1
\end{tcolorbox}
}

\definecolor{codegray}{gray}{0.9}
\newcommand{\cc}[1]{\colorbox{codegray}{\footnotesize\texttt{#1}}}

\newcommand{\toolname}{{\sc LeakPair}\xspace}
\newcommand{\nknown}{18\xspace}
\newcommand{\nunknown}{19\xspace}
\newcommand{\nsubjects}{37\xspace}
\newcommand{\coconut}{{\sc CoCoNuT}\xspace}
\newcommand{\footpatch}{{\sc FootPatch}\xspace}
\newcommand{\saver}{{\sc SAVER}\xspace}

\begin{document}

\title{\toolname: Proactive Repairing of \\ Memory Leaks in Single Page Web Applications}

\author{\IEEEauthorblockN{Arooba Shahoor}
	\IEEEauthorblockA{\textit{Kyungpook National University} \\
		Daegu, Republic of Korea \\
		arooba.shahoor@knu.ac.kr}
	\and
	\IEEEauthorblockN{Askar Yeltayuly Khamit}
	\IEEEauthorblockA{\textit{UNIST} \\
		Ulsan, Republic of Korea \\
		khamit.askar@unist.ac.kr}
	\and
	\IEEEauthorblockN{Jooyong Yi}
	\IEEEauthorblockA{\textit{UNIST} \\
		Ulsan, Republic of Korea \\
		jooyong@unist.ac.kr}
	\and
	\IEEEauthorblockN{Dongsun Kim$^{\dagger}$\thanks{$^{\dagger}$Corresponding author.}}
	\IEEEauthorblockA{\textit{Kyungpook National University} \\
		Daegu, Republic of Korea \\
		darkrsw@knu.ac.kr}
}


\maketitle

\begin{abstract}
Modern web applications often resort to application development frameworks such as React, Vue.js, and Angular. While the frameworks facilitate the development of web applications with several useful components, they are inevitably vulnerable to unmanaged memory consumption since the frameworks often produce Single Page Applications (SPAs).
Web applications can be alive for hours and days with behavior loops, in such cases, even a single memory leak in a SPA app can cause performance degradation on the client side.
However, recent debugging techniques for web applications still focus on memory leak detection, which requires manual tasks and produces imprecise results.

We propose \toolname, a technique to repair memory leaks in single page applications.
Given the insight that memory leaks are mostly non-functional bugs 
and fixing them might not change the behavior of an application,
the technique is designed to proactively generate patches to fix
memory leaks, without leak detection, which is often heavy and tedious.
To generate effective patches, \toolname follows the idea of pattern-based program repair
since the automated repair strategy shows successful results in many recent studies.
We evaluate the technique on more than 20 open-source projects without using explicit leak detection.
The patches generated by our technique are also submitted to the projects as pull requests.
The results show that \toolname can generate effective patches to reduce memory consumption that are acceptable to developers. In addition, we execute the test suites
given by the projects after applying the patches, and it turns out that
the patches do not cause any functionality breakage; this might imply that
\toolname can generate non-intrusive patches for memory leaks.
\end{abstract}





\section{Introduction}


\begin{figure}[t!]
    \subfloat[Event listener memory leak in Rooster JS.]{\includegraphics[width=\linewidth]{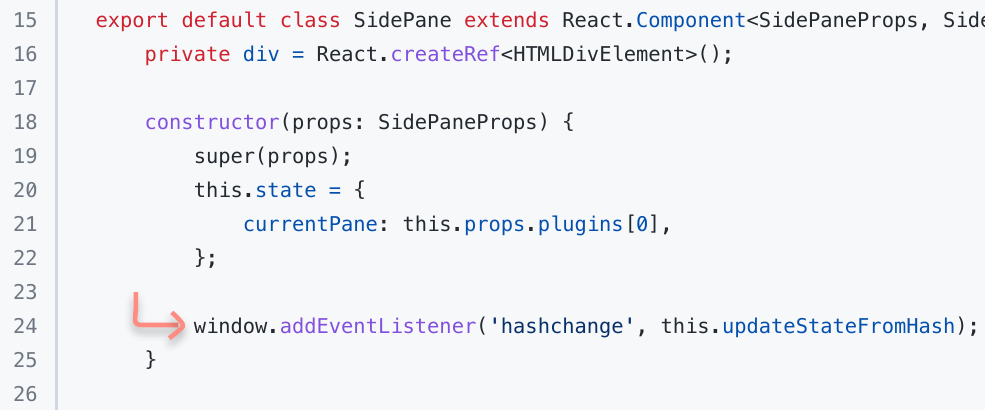}}
    \\
    \subfloat[Patch for the memory leak in (a).]{\includegraphics[width=\linewidth]{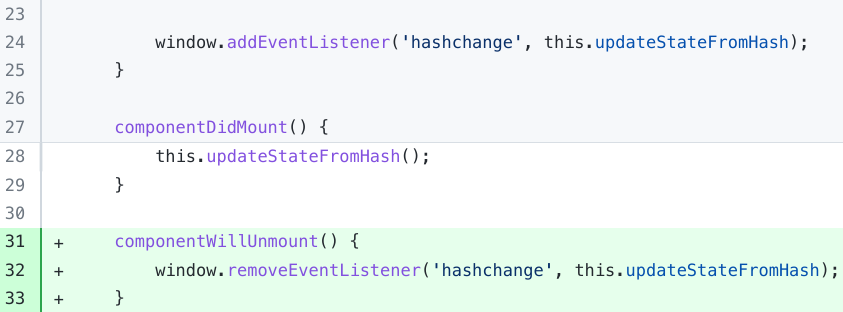}}
    \caption{Memory leak in Rooster JS~\cite{c3f2f0-microsoft2023Feb} and its corresponding patch.}
    \label{fig:leak:fix}
\end{figure}

Up until 2010, the website realm was mostly comprised of MPAs (Multiple Page Applications), where each page had to re-fetch and reload the entire webpage for each user request. The traditional MPA approach incurs a longer page switch time owing to the server round-trip for each request, and this delay increases with the size and complexity of the server APIs. The burgeoning usage of smartphones and mobile apps and the growing demands for swift and responsive web apps inspired the web development community to change how web pages were architected and rendered.

To address the responsiveness of web pages, the concept of Single Page Applications (SPAs) was first implemented by AngularJS, whereby rather than updating the entire webpage, only the data of the same page was updated~\cite{Lazarov2011-sb}. In SPAs, instead of re-fetching and loading entire pages from the server upon each request, just the data (usually in JSON format) can be retrieved asynchronously from the server and inserted dynamically into the application, thereby preventing page reloads on navigation and data fetch requests~\cite{Fink2014May}. Today, almost all contemporary social media apps make use of this architecture~\cite{bloomreach.com2023Jan}.

SPAs, however, are vulnerable to memory bloating due to their architecture in contrast to MPAs. Literally, SPAs maintain a single web page for a specific application, and every object should reside in a single page.
Therefore, SPAs inevitably rely on the garbage collectors of browsers to manage the memory space. Moreover, SPAs are highly likely to retain many loops
(i.e., navigate back to the previous page) and the loops can
rapidly add unnecessary objects that do not get garbage-collected due to unintentional reference. Such leaks might not be a problem in MPAs, where on each page navigation, the page refreshes, clearing all the heap. In SPA, however, such leaks can easily accumulate to several megabytes as a single page remains alive for several hours or even days.


Because such memory-leaking patterns are not syntactically or semantically invalid code, browsers run the program without throwing any errors, and they go unnoticed in functional testing as well~\cite{four-types-of-leaks}. Consider the syntactically and semantically correct code scenario in Figure~\ref{fig:leak:fix}(a) from Microsoft’s \cc{roosterjs} library~\cite{roosterjs}.
Based on the React framework, the class adds a listener for a \cc{hashchange} event (an event that is fired every time the part of the URL after the hash changes~\cite{hashchange}), to each new instance of the class, without ever removing the listener, even after the component unmounts from DOM. This created a memory leak in the application.

An important point to note in the above scenario is that if the listener handler was attached to a local element that does not have references to any other object, it would have been automatically cleaned up by the garbage collector (GC) once the class instance was destroyed. In the above case, however, the event is attached to the global (window) object, which the GC never cleans up, even after the instance is destroyed. A simple fix to this memory leak was applied by the project developers (Figure~\ref{fig:leak:fix}(b)) by explicitly removing the event in the component destructor function.

There have been a limited number of studies~\cite{9671473, meminsight, Lightweight, leakspot, bleak} on the problem of memory leak detection in the web domain. These studies focus on automating the detection of memory leaks, the most relevant and notable of which is BLeak~\cite{bleak}, which is an automated memory leak detection tool for client-side web applications. BLeak requires a scenario file written by the users to run the app in a loop in a headless browser and takes around 10 minutes to execute. The details of other studies will be presented later in the Related Work section.


We present \toolname, an approach to generating patches that repair memory leaks in SPAs.
Unlike typical automated program repair approaches, \toolname can be applied without requiring bug locations or relying on leak-detection techniques. It automatically detects code snippets that can potentially cause memory leaks and fixes them using non-intrusive (i.e., functionality-preserving) transformation rules we mined from existing code.

While test-driven program repair~\cite{nguyen_semfix:_2013,fixminer,tbar,6606626} (also known as
generate-and-validate repair~\cite{goues_automated_2019}) begins to work
once a bug is detected by test cases, proactive program repair first
applies patches to potential buggy locations. Then, a proactive approach
measures a difference of properties (such as memory consumption and
execution time) between before and after applying the patches. The
difference is provided as evidence of repair instead of validating patches
by test cases, which is done in test-driven program repair after generating patches.
Thus, proactive repair is a special kind of program repair approaches.



In summary, this paper contributes the following:
\begin{itemize}
    \item \toolname, a novel proactive approach to generating non-intrusive patches for fixing memory leaks in single page applications (SPAs).
    \item Four behavior-preserving fix patterns dedicated for repairing memory leaks in SPAs, which \toolname can leverage to generate non-intrusive patches.
    \item Empirical results of evaluating our approach on \nsubjects open-source projects, which show the effectiveness and non-intrusiveness when repairing memory leaks in SPAs.
\end{itemize}


\section{Background and Motivation}

\subsection{Single Page Web Applications (SPAs)}

This section compares Multiple Page Applications (MPAs) and Single Page
Applications (SPAs) and discusses why SPAs are vulnerable to memory leaks.

In MPA, the actions taken by the user on the webpage trigger HTTP requests to
the server; the server responds with a new page for each request, which means
a page refresh for each interaction. In addition, the user session and data
are persisted on the server; any time the session state or data is needed or
updated, the server needs to be queried, and the client (and the user) needs
to wait for the update to be completed on the server, resulting in poor app
responsiveness~\cite{Fink2014May}.


In contrast, SPA implements the majority of the logic for view generation on
the Client side. A single \cc{.html} file is loaded once, at the start of the
program load, which is the only full browser load throughout the app. The single
file contains multiple templates for different ‘views’, which are rendered on demand.
Upon a user query, data is fetched from the server, and a template is updated
with the date in real-time, without requiring a page reload. In addition,
SPA caches all the received data from the server so that the user is still
able to interact with the app in case of poor connection or connection loss,
and any new data can be synced once the connection improves/restores~\cite{Fink2014May}.


In SPA, the job of merging data with views moves from the server to the client.
The single HTML file (1) contains templates where data can be inserted and
(2) generates a new `view' that is equivalent to a new page in MPA. The logic of
merging the data with the right template, routing to the right view, and maintaining
the life cycle of a single view is accomplished via SPA frameworks such as Angular,
React, Vue etc. When a user navigates to a ``new page'', the SPA framework is simply
switching from one ``view'' to another~\cite{bloomreach.com2023Jan}.

\subsection{Garbage Collection and Memory Leaks in SPAs}

In MPAs, memory leakage may not be a critical issue since the web pages are
switched frequently and, as the browser switches to a new page, the memory
reserved by the previous page is reclaimed by the garbage collector.
Most modern web apps, however, are single-page apps that update the content
without switching the web page. This means that a single web page can be active
for several hours or even days~\cite{Lightweight}. When memory leakage in such
applications accumulates over time, it not only slows the program execution and
causes data processing latency but may also lead to program crashes and
incompatibility with other applications.


Several existing popular websites (including the libraries they use) suffer
from memory leakage that adversely affects the responsiveness of the browser.
Vilk and Berger~\cite{bleak} reported that more than 99 percent of Google
Chrome crashes on low-end Android phones are the result of memory issues. They
also identified more than 50 memory leaks in popular applications, including
JavaScript frameworks, and Google applications. Another leak detection
study~\cite{fuite} revealed public-facing SPAs leaking up to 186 MB per interaction.

Furthermore, as will be demonstrated in the next section, since such leaks are hard to diagnose, developers rather choose to invest their time and effort addressing more `apparent' application issues. Finally, oftentimes developers may wrongly attribute the lagging app behavior to user’s browser, internet connection, or even their systems.

\subsection{Non-intrusive repair without replicating actual memory leaks}

We figured out that it is challenging and non-trivial if a developer
tries to diagnose memory leaks in SPAs.
Unlike manually managed languages (such as C and C++), the JavaScript standard (ECMAScript), does not provide any interface for developers to monitor the memory usage of the app or manipulate the Garbage Collector, which makes diagnosing the leaking memory a cumbersome task for the developers~\cite{Conrod2023Feb}. Consider testimonials~\cite{MarcoPereira272023Feb,63661738-stackoverflow,angular27803} as well as the following comments from SPA developers on Github and StackOverflow regarding the obscure and evasive nature of memory leaks and their detection:
\myquote{I looked at the Chrome Dev Tools and taking heap snapshots to see if there is an increase in memory and it is apparent that there is when I see the memory shoot from 123MB to 200+MB after a few actions within the application. Now this is a good tool for determining whether there is a possible memory leak or not, but it's absolutely hard to read and understand, which doesn't help me determine where the issues lie~\cite{63813604-stackoverflow}.}
\vspace{2pt}
\myquote{This issue has been around for nearly 3 years now. (I usually don't like to start a message this way unless I tried something to fix the issue myself... Which I did here! and failed miserably as it seem quite complex to get to the bottom of it...\cite{angular20007}.}

In order to address memory leak issues, the root cause needs to be diagnosed first.
Although there have been automated techniques and approaches
to detect memory leaks in web applications~\cite{9671473, meminsight, Lightweight, leakspot, bleak}, these techniques have several limitations, including (1) dependency on the browser's heap snapshots, (2) non-trivial effort required for writing a test-driver script
and (3) imprecision.

\noindent
{\bf Non-intrusive patches}: Our intuition here is to
apply non-intrusive patches~\cite{nistorcaramel2015} to all potential memory leaks.
If the patches are non-intrusive (i.e., behavior-preserving),
it is not necessary to detect memory leaks before repairing them. As the patches do not change
the behavior of a target program, it is better to repair as many (potential) leaks as possible, which eventually improves the maintenance quality.
Such patches are unlikely to introduce new functional bugs and often easy to
understand. The tradeoff for developers is obvious: applying these patches are beneficial as they are simple and non-intrusive.
Avoiding the leak detection step is a huge advantage, as this step
is tedious and time-consuming due to the dynamic analysis involved.
A similar approach was used in \cite{nistorcaramel2015} to fix performance bugs.
However, ours is the first work using non-intrusive patches to fix memory leak issues, to the best of our knowledge.

\noindent
{\bf Pattern-based program repair}:
To fix the memory leak issues, we employ pattern-based program repair.
While we considered other types of program repair techniques as well, they were found to be less suitable for fixing memory leaks proactively.
Most existing APR techniques (e.g.\cite{weimer_automatically_2009,nguyen_semfix:_2013,tbar}) are test-driven, meaning that they require a test suite to drive the search for a patch, while we do not assume the existence of such a test suite.
Note that recent neural program repair techniques (e.g.\cite{zhu_syntax-guided_2021,lutellier_coconut_2020}) also require a test suite to validate the generated patches.
As will be shown in Section~\ref{sec:coconut}, the current neural program repair techniques such as \coconut~\cite{lutellier_coconut_2020} are not capable of fixing the memory leaks of SPAs.
The issue of the trustworthiness of the generated patches is also a concern for such techniques.

In comparison, we curate patch patterns that are likely to be non-intrusive and apply them to the potential memory-leak locations of the program.
Our pattern-based program repair can also be viewed as static-analysis-based repair similar to \footpatch~\cite{footpatch} and \saver~\cite{saver}, tools fixing the memory leaks of C/Java~\!\!\footnote{SAVER cannot handle Java programs.} programs --- we statically detect potential memory-leak locations and fix them.
These techniques typically involve substantial efforts of both tool developers and users to enable static analysis.
For example, \saver requires the semantic models for libraries to perform static analysis and fixing.
By contrast, our pattern-based approach does not involve any heavyweight analysis and can be readily applied to any SPA program.
The event-driven / object-oriented nature of SPAs makes it easy for the developers to assess the correctness of the patches --- a patch is often applied to an object destructor which is called implicitly when an event occurs.
As will be shown in Section~\ref{sec:acceptability}, the patches generated from \toolname are often accepted by real-world developers, demonstrating the practical value of our approach.


\section{\toolname}

\subsection{Overview}
Our approach, \toolname, consists of two steps: (1) fix pattern mining and (2)
memory leak repair using the fix patterns. \ul{In the first step}, we manually examined program patches or pull requests addressing memory leaks, together with commit messages, code reviews available in open source projects, and Q\&A posts. After identifying common and recurring fix patterns from the patches, we implement an edit script for each pattern, which can generate non-intrusive patches. \ul{In the second step}, we scan a target project (i.e., SPAs) to apply our fix patterns. Each fix pattern can naturally specify which data or object types are associated with it. A corresponding edit script can then be applied accordingly. Each pattern changes all locations, where applicable, in the target project.

\subsection{Mining fix patterns for SPA memory leaks}
\label{sec:mining}

Since our goal is to identify recurring and common patterns of memory leaks and
their corresponding patches in SPAs, we first collected the most common leaks available
publicly by using specific keywords such as `leak' and `React'. Our search targets were \cc{GitHub.com} and \cc{stackoverflow.com}. Then, we carefully extracted common patterns of leaks and their corresponding patches. Obviously, this is a manual task and is time-consuming. Nonetheless, numerous previous studies ~\cite{6606626,Vejovis,nistorcaramel2015,8330202,3213871,fixminer,tbar} have demonstrated that this strategy is effective and useful, as we can reuse the fix patterns many times once they have been identified.


We used the following search process to collect issues and discussions relevant to memory leaks: (1) For Stack Overflow, we searched through 1,000 posts whose titles, comments, or discussions contain keywords such as `leak', `memory usage' or `memory leak' and that explicitly pertained to JavaScript applications, (2) For GitHub, we searched through 1,000 commits, PRs, issues, and discussions containing any of the above keywords, along with the labels `React' or  `Angular' (being the most commonly used frameworks for SPA development).

After investigating the search results, we collect leak patterns as per the following procedures:
(1) We selected common memory leaks reported at least five times across
\cc{GitHub.com} and \cc{stackoverflow.com}, (2) the leaks should be acknowledged as valid, by at least two developers, (3) we further narrowed down the
leaks, which can be reproduced and tested locally, and (4) four leak patterns were selected, which are applicable to SPAs.

For each leak pattern identified in the previous step, we selected fix patterns by looking at their original answers (for StackOverflow) or discussions (for GitHub). For each leak type, we extracted, as fix patterns, the common fix suggestions in Stack Overflow that are accepted as the answer in at least two separate posts. From the leak patterns found in GitHub commits, we selected the patches that were approved and merged in at least two separate projects.
Among the above-selected fix patterns, we further filtered the patterns based on their applicability to SPA projects.

All identified fix patterns are supported by examining actual memory footprint changes. We compare the memory footprints of revisions before and after applying the patches. If there are no differences between \cc{before} and \cc{after} memory footprints, we discard the fix patterns. We examined the memory footprints of patches applied to SPAs using MemLab~\cite{memlab}.

\subsection{Fix patterns}
\label{sec:patterns}


As already discussed in Section 2.2, the general root cause of memory leaks in SPA is an unused object that lingers in memory due to some unwanted reference that was not explicitly cleared by the developers. Hence, the fix for such leaks generally involves cleaning up any unwanted references to objects that have the potential to be retained in memory. In the SPA domain, this needs to be done when a \cc{component} unmounts from the DOM (in the component destructor).

Following the procedure in Section ~\ref{sec:mining}, we identified four fix patterns for generating non-intrusive patches for repairing memory leaks in SPAs:

\noindent
\underline{\bf FP1. Unreleased Subscription.}
In reactive JavaScript (RxJS), an \cc{Observable} is a lazily evaluated computation that can synchronously or asynchronously return zero to (potentially) infinite values from the time it is invoked (subscribed)~\cite{rxjs}. This indicates that they can keep outputting values even after the component is destroyed/unmounted,  unless we explicitly tell them to stop. This means each time the component containing that subscription is rendered, a new observable is created in addition to the old one, because we never explicitly unsubscribed from the previous one. The stale data keeps getting piled up, never getting garbage collected, creating a memory leak.

In practice, developers may not always be able to figure out whether the observable they are subscribing to, is finite or infinite, and in these cases, it is best to explicitly \cc{unsubscribe} when the component unmounts/destroys, just to be safe. This ensures that the Subscription is closed (if it was not already) and that proper cleanup is carried out. Nothing else will happen if it was previously closed.

\noindent
{\bf Fix:} The \cc{takeUntil()} operator allows a notified \cc{Observable} to emit values until a value is emitted from another Observable \cite{takeuntil}, i.e., the \cc{takeUntil()} operator completes the stream it is attached to, when an Observable provided to itself, emits a value. Thus, if we provide \cc{observer2} (see pseudo-code below) as input to the \cc{takeUntil()} operator, and in the destructor we make \cc{observer2} emit a value (using the \cc{next()} and \cc{complete()} methods), that will clear the subscription and thus prevent the memory leak\footnote{\url{https://github.com/blackbaud/skyux/pull/376/files}}.
\\
\fixpattern{
\begin{lstlisting}[language=java,linewidth={\linewidth},basicstyle=\scriptsize\ttfamily]
(@\textcolor{red}{-}@) observer1.subscribe(() => {...})
(@\textcolor{codegreen}{+}@) observer1.pipe(takeUntil(observer2)).subscribe(() => {...})
...
(@\textcolor{codegreen}{+}@) destructorMethodDeclaration() {
(@\phantom{x}\hspace{6.7ex}@) ...
(@\textcolor{codegreen}{+}\phantom{x}\hspace{6.7ex}@)observer2.next()
(@\textcolor{codegreen}{+}\phantom{x}\hspace{6.7ex}@)observer2.complete()
(@\textcolor{codegreen}{+}@) }
\end{lstlisting}}
\noindent


\noindent
\underline{\bf FP2. Unremoved Event Listener.}
The notion of {\it retaining paths} is critical for finding the root cause of a
memory leak. A retaining path is a chain of objects that prevents the garbage collection of the leaking object. The chain starts at a root object, such as the
global object of the {\bf main window}. The chain ends at the leaking object.

Active event listeners will prevent all variables captured in their scope from being garbage-collected. Once added, the event listener will remain in effect until (1) it is explicitly removed with \cc{removeEventListener()} or (2) the associated DOM element is removed.

\noindent
{\bf Fix:}
Unregistering the event listener once the SPA component unmounts/destroys, by creating a reference pointing to the \cc{listenerHandler} (see pseudo-code below) and passing it to \cc{removeEventListener()} method\footnote{\url{https://github.com/microsoft/roosterjs/pull/921/files}}.
\\
\fixpattern{
\begin{lstlisting}[language=java,linewidth={\linewidth},basicstyle=\scriptsize\ttfamily]
function listenerHandler() {
(@\phantom{x}\hspace{6.7ex}@) ...
}
...
eventTarget.addEventListener(eventType, listenerHandler , options)
...
(@\textcolor{codegreen}{+}@) destructorMethodDeclaration() {
(@\phantom{x}\hspace{6.7ex}@) ...
(@\textcolor{codegreen}{+}\phantom{x}\hspace{6.7ex}@)eventTarget.removeEventListener(eventType, listenerHandler, options)
(@\textcolor{codegreen}{+}@) }
\end{lstlisting}}

\noindent
\underline{\bf FP3a. Uncleared Timers: \cc{setTimeOut}.}
The \cc{setTimeout()} method executes a function or specified piece of code once the specified timeout value is reached. When any object is tied to a timer callback, it will not be released until the timeout happens. In certain scenarios, the program’s logic requires the timer to reset itself; this causes it to run forever, thereby retaining the references of all the enclosing objects and disallowing the garbage collector to remove the memory. Even if the developers explicitly clear the \cc{setTimeout()} in code conditionally, there is no guarantee it also caters for situations where the user navigates away after the \cc{setTimeout()} is triggered but before the specified timeout value is reached.

\noindent
{\bf Fix:}
Because each \cc{setTimeout()} has its own memory reference, we must clear each one individually, using the \cc{clearTimeout()} method, passing it the ID returned from the \cc{setTimeout()} call (which uniquely identifies each \cc{setTimeout()} reference). The patch involves clearing the timeout method just before the component is about to unmount from DOM i.e in the component destructor\footnote{\url{https://github.com/MTES-MCT/monitorfish/pull/953/commits/1dc01c0d82261bf05277366d954fa5d912632091}}.
\\
\fixpattern{
\begin{lstlisting}[language=java,linewidth={\linewidth},basicstyle=\scriptsize\ttfamily]
(@\textcolor{red}{-}@) setTimeOut(() => {...})
(@\textcolor{codegreen}{+}@) timeOutID = setTimeOut(() => {...})
...
(@\textcolor{codegreen}{+}@) destructorMethodDeclaration() {
         ...
(@\textcolor{codegreen}{+}\phantom{x}\hspace{6.7ex}@)clearTimeOut(timeOutID)
(@\textcolor{codegreen}{+}@) }
\end{lstlisting}}


\noindent
\underline{\bf FP3b. Uncleared Timers: \cc{setInterval}.}
The \cc{setInterval()} method repeatedly calls a function or executes a code snippet, with a fixed time delay between each call. Even after the component is unmounted from the DOM, the setInterval timer will keep on ticking (unless we explicitly clear the interval in the code), trying to update the state of a component that’s effectively gone, thereby causing memory leakage~\cite{Codecademy}. Even if the developers clear these interval functions in the code on some condition, there is no guarantee that the clearing method will get a chance to execute before the user navigates away.

\noindent
{\bf Fix:}
Each interval has a separate reference in memory, so we need to clear each individually, using the returned ID from the \cc{setInterval()} method call, which uniquely identifies the interval method call. The patch involves clearing the timer just before the component is about to be destroyed i.e., in the component destructor\footnote{\url{https://github.com/MTES-MCT/monitorfish/pull/953/files}}.
\\
\fixpattern{
\begin{lstlisting}[language=java,linewidth={\linewidth},basicstyle=\scriptsize\ttfamily]
(@\textcolor{red}{-}@) setInterval(() => {...})
(@\textcolor{codegreen}{+}@) intervalID = setInterval(() => {...})
...
(@\textcolor{codegreen}{+}@) destructorMethodDeclaration() {
         ...
(@\textcolor{codegreen}{+}\phantom{x}\hspace{6.7ex}@)clearInterval(intervalID)
(@\textcolor{codegreen}{+}@) }
\end{lstlisting}}


\noindent
\underline{\bf FP4. Uncancelled Animation Frame Requests.}\\
The \cc{requestAnimationFrame()} Web API method helps determine the count of frames per second to allocate an animation, and execute the provided callback to perform that animation, before the actual screen loads~\cite{requestAnimationFrame}. Since it is used for creating animations on web pages, these are usually called recursively, which again leads to the risk of their execution post component destruction, retaining all objects in its callback function, even after they are no longer needed.

\noindent
{\bf Fix:}
Similar to timers, each \cc{requestAnimationFrame()} call also returns an ID unique to that specific request, that we can use to ensure the request is cancelled just before the component destroys\footnote{\url{https://github.com/carbon-design-system/carbon-addons-iot-react/pull/2119/files}}.
\\
\fixpattern{
\begin{lstlisting}[language=java,linewidth={\linewidth},basicstyle=\scriptsize\ttfamily]
(@\textcolor{red}{-}@) requestAnimationFrame(() => {...})
(@\textcolor{codegreen}{+}@) let requestID = requestAnimationFrame(() => {...})

(@\textcolor{codegreen}{+}@) destructorMethodDeclaration() {
         ...
(@\textcolor{codegreen}{+}\phantom{x}\hspace{6.7ex}@)cancelAnimationFrame(requestID)
(@\textcolor{codegreen}{+}@) }
\end{lstlisting}}


\subsubsection{Edit script}
\label{sec:editscript}

For each individual fix pattern, we create a corresponding edit script to
actually generate patches for potential memory leaks. An edit script is another
program that parses the target program and locates potential leaking objects,
where we apply the fix pattern. Each edit script has two components: (1) a potential leak object locator and (2) patch writer.
Creating edit scripts is a common procedure
when applying a pattern-based program repair technique~\cite{6606626,fixminer,avatar,tbar,nistorcaramel2015}.
Therefore, we implement the scripts for our tool, which are available in our replication package~\cite{figshare}.

\subsubsection{Coverage of the patterns}
The four fix patterns cover most of the fixed memory leaks we have examined.
Following the procedures described in Section~\ref{sec:mining},
we identified 124 and 65 memory leak bugs in SPAs based on React and Angular,
respectively, as a result.
 These bugs have been confirmed and fixed by the developers of the SPAs.
Our four fix patterns can also fix 102 out of 124 (82\%) and 57 out of 65 (88\%)
already-known React and Angular-related memory leak bugs, respectively.
The full list of known memory leak bugs examined is available in our
replication pacakge~\cite{figshare}.


\subsection{Applying fix patterns}
\label{sec:patternapplication}

As the second step, \toolname applies the fix patterns extracted in the first step (Section~\ref{sec:mining}). Basically, we assume that one can apply \toolname to
the whole project by scanning the source code tree of the project, which implies that the edit scripts explained in Section~\ref{sec:editscript} are executed for each file.
Specifically, it follows the following procedure.


\begin{itemize}
    \item {\bf Parsing and Detecting:} \toolname makes use of the Babel compiler \cite{babel-parser} in conjunction with Facebook’s \textit{jscodeshift} ~\cite{jscodeshift} to traverse through the JS file (in the case of a single file path) or all JavaScript files from the root of the given project path.
For each file, it extracts the AST by leveraging the Babel compiler. During the AST traversal, \toolname detects Angular and React components (Vue is not supported currently) by matching their syntax definition. Once a component from these frameworks is identified, it detects whether the component implements any of the four memory leak patterns by traversing the AST, visiting each node, and matching the patterns illustrated in Section~\ref{sec:patterns}.
    \item {\bf Creating Patches:} If a leak pattern is matched, it tracks the file name as well as how many objects are leaking due to that leak type, i.e., are following the same pattern, in that specific component. It then generates and adds the fix (which is predominantly clearing the leaking objects in the component destructor) in the AST. After the patch is successfully applied, it updates the count of potentially leaking objects for that leak type, in the project/file. Finally, it then converts the AST back to source code by leveraging the Recast~\cite{recast} library.
    \item {\bf Repeating and Reporting:} \toolname repeats this process for all the files if a project path was specified; otherwise, the processing completes there. At the end of the execution, it prints out the repaired file name(s) as well as the total count of each leak type.
\end{itemize}

\subsection{Non-intrusive patch generation}
As \toolname aims at proactively generating non-intrusive patches for memory
leaks in SPAs, we apply the following procedures in addition to the
steps of standard pattern-based program repair techniques~\cite{6606626,tbar}:
\begin{itemize}
  \item {\bf Localizing without test cases:} Since \toolname proactively
  generates patches for memory leaks in SPAs, it does not rely on external fault localization techniques usually based on test suites. Instead, our approach scans specific objects in the source code.
  For example, FP1 detects all \cc{Observable} objects in the target SPA.
  \item {\bf Avoiding redundant fix:} Among the detected target objects,
  some of them are correctly used and memory leaks are prevented, where
  \toolname does not need to generate corresponding patches. Our approach
  carefully scans the target SPA once again to figure out whether there is
  any cleanup code for the specific object for each pattern. \toolname generates a patch by using the pattern only when there is no cleanup code to avoid redundant patches, which can unnecessarily bloat the source code.
  \item {\bf Checking non-intrusiveness:} For each generated patch,
  \toolname examines whether the patch breaks any functionality. As the
  regression test suites are often available for a target SPA, our approach
  runs the suites to find any behavior changes. Although test cases
  may not guarantee complete behavior integrity, the test results may
  show the correctness of key functionalities for the target SPA.
\end{itemize}


\section{Evaluation}
\label{sec:eval}

\begin{table}[!t]
  \centering
	\scriptsize
	\setlength\tabcolsep{2pt}
	\caption{Subjects with unknown memory leaks.}
	\label{tab:subject:unknown}
	\resizebox{1.00\linewidth}{!}
 {%
\begin{threeparttable}
    \begin{tabular}{c|c|c|c|c}
    \toprule
    \textbf{ID} & \textbf{Program}                    & \textbf{Type} & \textbf{SPA Framework} & \textbf{Commit Hash} \\ \midrule
    U1               & react-zoom-pan-pinch~\cite{prc52023Feb}        & Library               & React                  & fdc030                           \\
    U2               & Angular Extentions Elements~\cite{angular-components}  & Library               & Angular                & d9a4e4                           \\
    U3               & Evergreen~\cite{evergreen} & Framework             & React                  & 82c3a8                         \\
    U4               & ngx-datatable~\cite{ngx-datatable}               & Library               & Angular                & 6184c9                          \\
    U5               & react-multi-carousel~\cite{react-multi-carousel}         & Library               & React                  & 525793                           \\
    U6               & codetekt (Frontend)~\cite{angular-ui}            & Website               & Angular                & 7b8289                           \\
    U7               & skbkontur/retail-ui~\cite{retail-ui}          & Framework             & React                  & 32f3cf                         \\
    U8               & Aam Digital~\cite{Aam-Digital2023Feb}                  & Web app               & Angular                & 304ff9                          \\
    U9               & Replay's DevTools~\cite{replayio2023Feb}            & Library               & React                  & 24d10f                         \\
    U10              & ngx-bootstrap~\cite{ngx-bootstrap}               & Framework             & Angular                & 663c70 \\
    \bottomrule
    \end{tabular}
    {The full list of subjects used for our experiment is available in the replication package~\cite{figshare}.}
\end{threeparttable}
}
\end{table}

\begin{table}[!t]
  \centering
	\scriptsize
	\setlength\tabcolsep{2pt}
	\caption{Subjects with known memory leaks.}
	\label{tab:subject:known}
	\resizebox{1.00\linewidth}{!}
 {%
\begin{threeparttable}
\begin{tabular}{c|c|c|c|c}
  \toprule
\textbf{ID} & \textbf{Program}                   & \textbf{Type} & \textbf{SPA Framework} & \textbf{Commit Hash} \\\midrule
K1               & react-zoom-pan-pinch~\cite{prc52023Feb} & Library               & React                  & 6e35b3                           \\
K2               & Fundamental Library for Angular~\cite{fundamental}     & Library               & Angular                & be9629                            \\
K3               & react-multi-carousel~\cite{react-multi-carousel} & Library               & React                  & 5d252d                           \\
K4               & Angular Components~\cite{angular-components}  & Framework             & Angular                & 1bbb29                           \\
K5               & Material UI~\cite{mui2023Feb}     & Framework             & React                  & e92b1c                           \\
K6               & Angular Components documentation~\cite{material.angular.io} & Website               & Angular                & e8cb0d                           \\
K7               & Rooster~\cite{roosterjs} & Library               & React                  & c3f2f0                           \\
K8               & Octant~\cite{octant}        & Framework             & Angular                & b079ad                           \\
K9               & Evergreen~\cite{evergreen}           & Framework             & React                  & a716f4                           \\
K10              & Transloco~\cite{transloco}           & Library               & Angular                & 2338a0\\
\bottomrule
\end{tabular}
{The full list of subjects used for our experiment is available in the replication package~\cite{figshare}.}
\end{threeparttable}
}%
\end{table}

\subsection{Research Questions}
\label{sec:rqs}

Our experiments investigate the following research questions:

\begin{enumerate}
  \item {\bf RQ1. (Effectiveness)} How effective is the tool at minimizing/eliminating memory leaks?
  \item {\bf RQ2. (Acceptability)} How useful are generated patches, as perceived by developers?
  \item {\bf RQ3. (Non-intrusiveness)} What is the impact of our tool on test suite execution results? 
\end{enumerate}

The first research question is designed to assess the amount of memory
reduction when applying \toolname to SPAs. For this RQ, we collected a set of known memory leaks and another set of unknown leaks in open source projects as experiment subjects. We apply our tool to the subjects and examine their memory footprints before and after repair.

While RQ1 assesses the effectiveness, RQ2 focuses on whether the patches generated by \toolname can be accepted by the developers of the open source projects. As the unknown leaks used in the experiments for RQ1 are in fact new defects, we report them as new pull requests and see whether they are merged or accepted.

As \toolname is designed to generate non-intrusive patches, it is necessary
to assess whether the patches make no changes to the functionality of the target
subjects  or cause compilation errors. Therefore, we designed RQ3 to
assess non-intrusiveness. Our experiments for this RQ try to
compile the subject programs used in the previous RQs and run the test cases
already given for the programs.

\subsection{Experiement Setup}
\label{sec:setup}

We use the following experiment design to answer the research questions
 described in Section~\ref{sec:rqs}.

\begin{table*}[!ht]
  \centering
	\scriptsize
	\setlength\tabcolsep{2pt}
	\caption{Memory consumption results before and after applying \toolname to the subjects in Table~\ref{tab:subject:unknown}.}
	\label{tab:result:unknown}
	\resizebox{\linewidth}{!}
 {%
\begin{threeparttable}
\begin{tabular}{c|c||c|c|c||c|c|c}
\toprule
\multirow{2}{*}{\bf ID} & \multirow{2}{*}{\bf Leak Patterns} &
{\bf Leaked Objects} & {\bf Leaked Objects} & {\bf Leak Object} & {\bf Heap Size Before} & {\bf Heap Size After} & {\bf Total Heap} \\
 & & {\bf Before applying \toolname} & {\bf After applying \toolname} & {\bf Reduction} & {\bf applying \toolname} & {\bf applying \toolname} & {\bf Size Reduction} \\ \midrule
U1 & FP3 & 4 clusters & 3 clusters & 1 cluster$\ast\ast$ & 47.9 MB & 43.7 MB & 5.2 MB (10.8\%)$\ast\ast$\\
U2 & FP1, FP2 & 13 clusters & 10 clusters & 3  clusters$\ast\ast$ & 17.4 MB & 16.7 MB & 0.7 MB (4\%)$\ast\ast$\\
U3 & FP4 & 5 clusters & 3 clusters & 2 clusters$\ast\ast$ & 35.5 MB & 29.2 MB & 6.3 MB (17.7\%)$\ast\ast$\\
U4 & FP3, FP4 & 8 clusters & 7 clusters & 1 cluster$\ast$ & 83.9 MB & 66.9 MB & 15.0 MB (17.8\%)$\ast\ast$\\
U5 & FP3 & 5 clusters & 4 clusters & 1 cluster$\ast\ast$ & 27.9 MB & 23.5 MB & 3.4 MB (12.1\%)$\ast$\\
U6 & FP1, FP3 & 12 clusters & 9 clusters & 3 clusters$\ast\ast$ & 65.1 MB & 63.6 MB & 1.5 MB (2.3\%)$\ast\ast$\\
U7 & FP3 & 7 clusters & 5 clusters & 2 clusters$\ast\ast$ & 105 MB & 100 MB & 5.0 MB (4.7\%)$\ast$\\
U8 & FP1 & 2 clusters (95,352 objects) & 2 clusters (73,951 objects) & 20,000+ objects$\ast\ast$ & 318.9 MB & 264.3 MB & 57.6 MB (18\%)$\ast\ast$\\
U9 & FP3 & 5 clusters & 3 clusters & 2 clusters$\ast\ast$ & 27 MB & 26.4 MB & 0.6 MB (2.2\%)\\
U10 & FP1 & 6 clusters & 4 clusters & 2 clusters$\ast\ast$ & 101.7 MB & 99.7 MB & 2.0 MB (1.9\%)$\ast\ast$\\\bottomrule
\end{tabular}
{$\ast$: p-value $<$ 0.05, $\ast\ast$: p-value $<$ 0.01. The full list of subjects used for our experiment is available in the replication package~\cite{figshare}.}
\end{threeparttable}
}
\end{table*}

\begin{table*}[!ht]
  \centering
	\scriptsize
	\setlength\tabcolsep{2pt}
	\caption{Memory consumption results before and after applying \toolname to the subjects in Table~\ref{tab:subject:known}.}
	\label{tab:result:known}
	\resizebox{\linewidth}{!}
 {%
\begin{threeparttable}
\begin{tabular}{c|c||c|c|c||c|c|c}
\toprule
\multirow{2}{*}{\bf ID} & \multirow{2}{*}{\bf Leak Patterns} &
{\bf Leaked Objects} & {\bf Leaked Objects} & {\bf Leak Object} & {\bf Heap Size Before} & {\bf Heapsize After} & {\bf Total Heap}\\
 & & {\bf Before applying \toolname} & {\bf After applying \toolname} & {\bf Reduction} & {\bf applying \toolname} & {\bf applying \toolname} & {\bf Size Reduction}\\ \midrule
K1 & FP2 & 3.9 clusters & 3.1 clusters & 0.8 clusters$\ast\ast$ & 28.19 MB & 28.09 MB & 0.1 MB (0.3\%)$\ast$\\
K2 & FP1 & 1 cluster (70.7 objects) & 1 cluster (66.9 objects) & 3.8 objects & 53.4 MB & 52.1 MB & 1.3 MB (2.4\%)$\ast\ast$\\
K3 & FP3 & 5 clusters & 4 cluster & 1 cluster$\ast$ & 17.19 MB & 16.87 MB & 0.32 MB (1.8\%)\\
K4 & FP1 & 5.6 clusters & 5.3 cluster & 0.3 cluster & 13.08 MB & 12.98 MB & 0.1 MB (0.7\%)$\ast$\\
K5 & FP3 & 13.8 clusters & 13.7 clusters & 0.1 clusters & 13 MB & 12.9 MB & 0.1 MB (0.7\%)\\
K6 & FP1 & 2 clusters (245.6 objects) & 2 clusters (154.5 objects) & 91.1 objects$\ast\ast$ & 11.8 MB & 11.8 MB & 0.0 MB (0\%)\\
K7 & FP2 & 4 clusters (4370.8 objects) & 4 clusters (4295.2 objects) & 75.6 objects$\ast$  & 10.9 MB & 10.87 MB & 0.03 MB (0.2\%)\\
K8 & FP1 & 16.4 clusters & 15.6 clusters & 0.8 clusters & 61.08 MB & 60.88 MB & 0.9 MB (1.4\%)\\
K9 & FP3 & 4 clusters (3923.9 objects) & 4 clusters (3662.7 objects) & 261.2 objects & 27.5 MB & 27.3 MB & 0.2 MB (0.7\%)$\ast$\\
K10 & FP1 & 1 cluster & 0 cluster & 1 cluster$\ast\ast$ & 9.2 MB & 9.2 MB & 0.0 MB (0\%)
\\\bottomrule
\end{tabular}
{$\ast$: p-value $<$ 0.05, $\ast\ast$: p-value $<$ 0.01. The full list of subjects used for our experiment is available in the replication package~\cite{figshare}.}
\end{threeparttable}
}
\end{table*}

\subsubsection{Subjects}
\label{sec:subjects}

To assess the effectiveness of our tool, we collected SPAs based on the
following criteria:
\begin{itemize}
  \item {\bf Maintained.} We chose projects that are still being maintained and whose last update was less than a year ago. Archived projects were not considered.
  \item {\bf Number of contributors.} Projects with at least 10 contributors were selected. Personal projects were not taken into account.
  \item {\bf Number of commits.} The selected projects have at least 100 commits on their GitHub repository.
  \item {\bf Popularity.} Projects with at least 10 stargazers, watchers, or forks were selected.
  \item {\bf Framework.} The selected projects should use either React or Angular as their base framework, as our target is SPAs.
\end{itemize}

Based on the above criteria, we collected a set of projects with {\it unknown} new
memory leaks and another set of projects with already
{\it known} leaks (i.e., those fixed by the developers).
The projects with already known leaks are necessary to show whether
our tool can reproduce the patches generated by the developers of the projects.
Other projects are collected to assess the effectiveness of our tool in discovering
and repairing new and unknown memory leak patterns.

As a result, \nsubjects projects are selected as the subjects for our
experiments to assess \toolname;  \nunknown projects have unknown new memory leaks while
\nknown projects out of them have already
known memory leaks.
In this paper, we focus on and report only on the results of 10 unknown and 10 known subjects
due to space limitations.
Tables~\ref{tab:subject:unknown} and~\ref{tab:subject:known} list 10 unknown and 10 known subjects, respectively, out of our \nsubjects subjects; the complete results of experiments for the entire set of subjects are available in our replication package~\cite{figshare}.

\subsubsection{Repairing memory leaks}
\label{sec:setup:rq1}

To answer RQ1, our first experiment applies our tool to the subjects described
in Section~\ref{sec:subjects}.
We run \toolname on the root of each subject so that it scans the project
directories and identifies JavaScript files. For each source code file,
the tool tries to change the file by applying each pattern.
Our tool addresses all locations if applicable.

After applying \toolname, we then measure the memory footprints.
Because we need to run the target subject to determine memory consumption, we created a scenario file for each subject. Using scenario files is a common procedure when measuring the memory consumption of web applications.
For example, BLeak~\cite{bleak} and MemLab~\cite{memlab}, the most recent techniques to detect memory leaks, require scenario files to run the target web applications.
The scenario files used for each subject are available in our replication package~\cite{figshare}.


To compare the memory consumption, we
compute the memory footprints before and after applying
\toolname. For each subject, the corresponding scenario file is
executed 10 times with \cc{loop=10}
(i.e., 10 $\times$ 10 times in total for each subject)
since a single loop may not accurately reveal
the memory consumption. We then collect memory consumption in megabytes (MB)
and the number of object clusters~\cite{memlab}, where a cluster is the collection of all retainer paths
for all the leaking objects due to a single leak origin.
Applying the Mann-Whitney U test~\cite{mann_test_1947},
we compute the statistical significance of the differences
between values before and after patches.
Note that this is not a stage of \toolname; rather, this is only
for the evaluation.


\subsubsection{Reporting generated patches}
\label{sec:setup:rq2}

As the unknown memory leaks are basically newly found bugs,
it is able to report the leaks to the repositories of the subjects.
For each patch generated by our tool, we create
a pull request with the patch and memory footprints before and after
applying the tool. The outcome of the reported pull requests
can be \cc{Agreed}, \cc{Disagreed}, or \cc{Ignored}. We count
each type of outcome to answer RQ2.

\subsubsection{Running test cases on patches}
\label{sec:setup:rq3}

To figure out whether the patches generated by \toolname break the functionality of the subjects, we execute
the test cases available in the subjects and count the number of passing and failing cases. As most of the popular open source projects maintain (regression) test suites, we simply run the test cases included in the subjects. Many subjects use test automation frameworks; in that case, we resort to those frameworks; otherwise, we follow the instructions available in the contribution guide for each subject. We also compare the number of passing/failing test cases before and after applying \toolname. The results of this experiment can answer RQ3.


\section{Results}
This section presents and analyzes the results of
experiments to answer the research questions
described in Section~\ref{sec:eval}.

\begin{figure*}[ht!]
  \centering
  {%
    \subfloat[U1]{\includegraphics[width=0.20\linewidth]{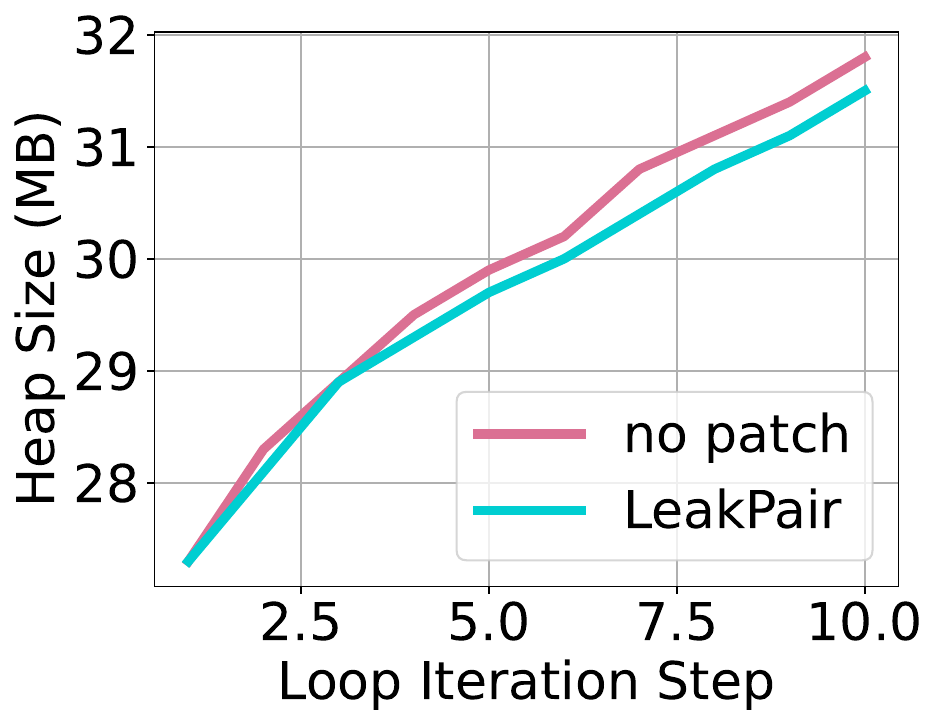}}
    \subfloat[U2]{\includegraphics[width=0.20\linewidth]{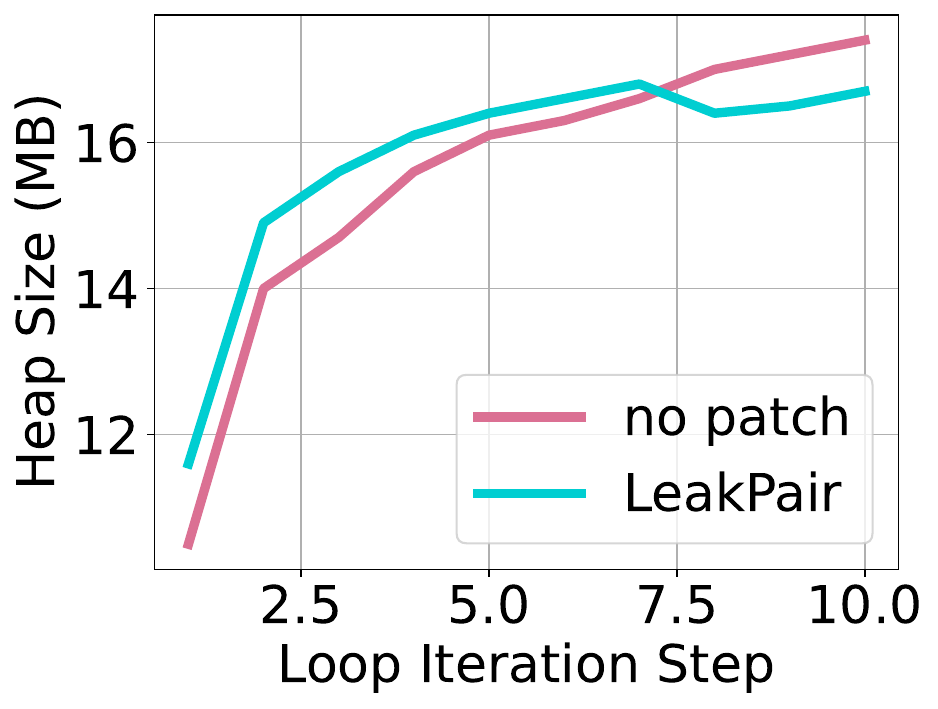}}
    \subfloat[U3]{\includegraphics[width=0.20\linewidth]{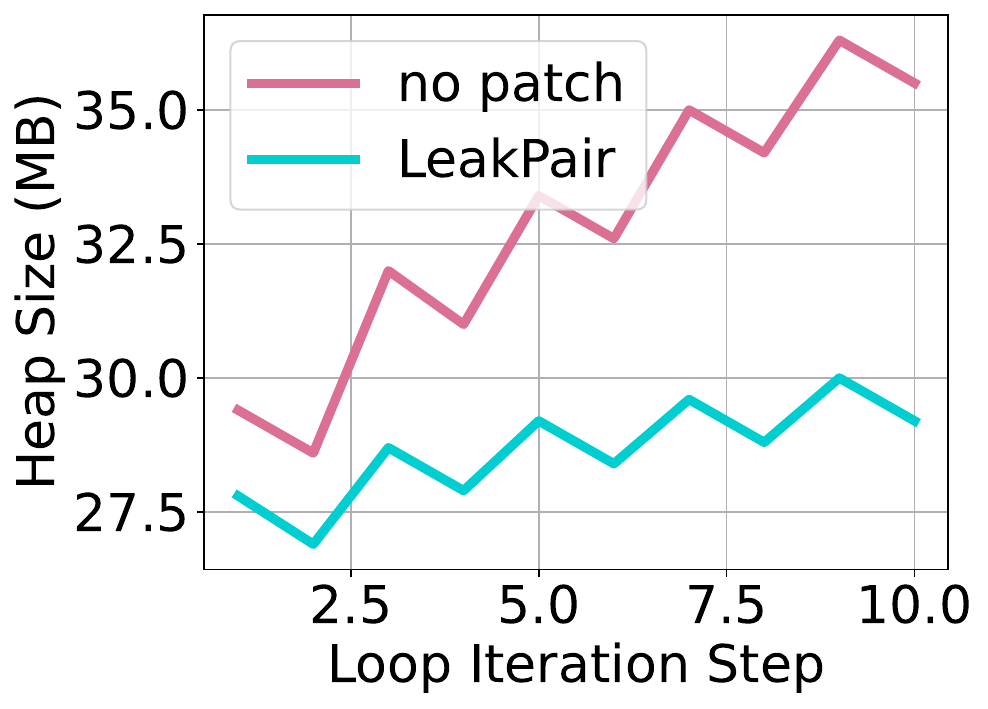}}
    \subfloat[U4]{\includegraphics[width=0.20\linewidth]{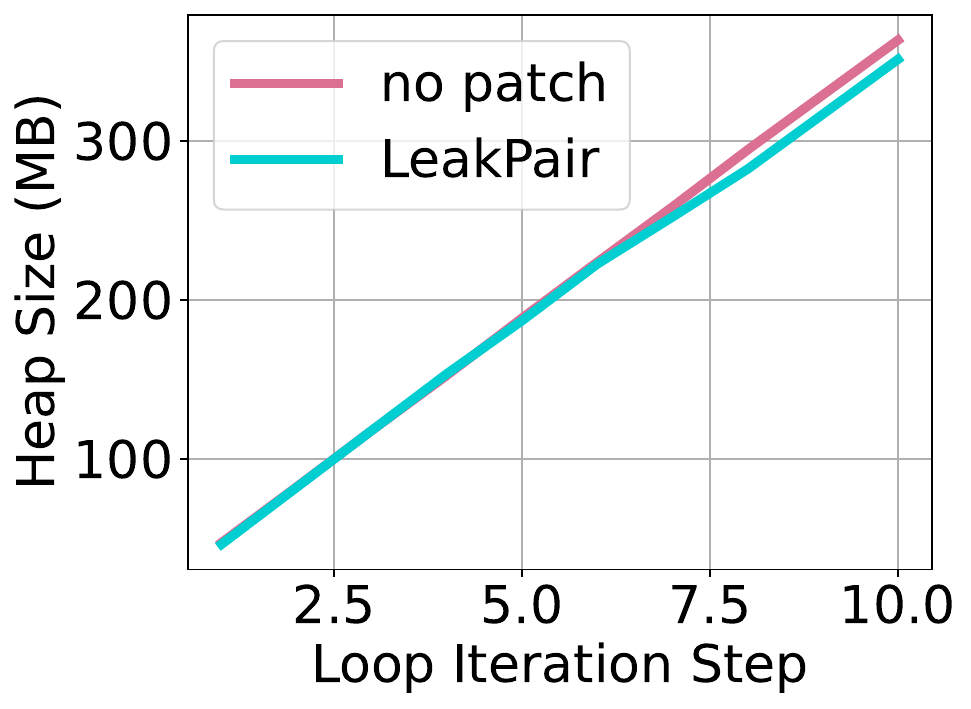}}
    \subfloat[U5]{\includegraphics[width=0.20\linewidth]{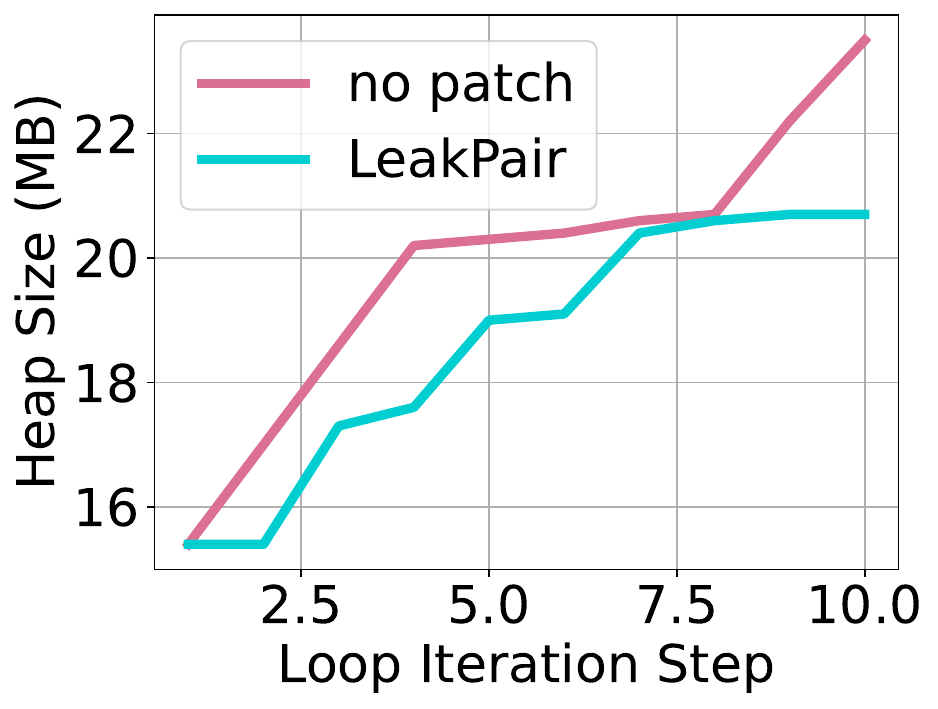}}
    \\
    \subfloat[U6]{\includegraphics[width=0.20\linewidth]{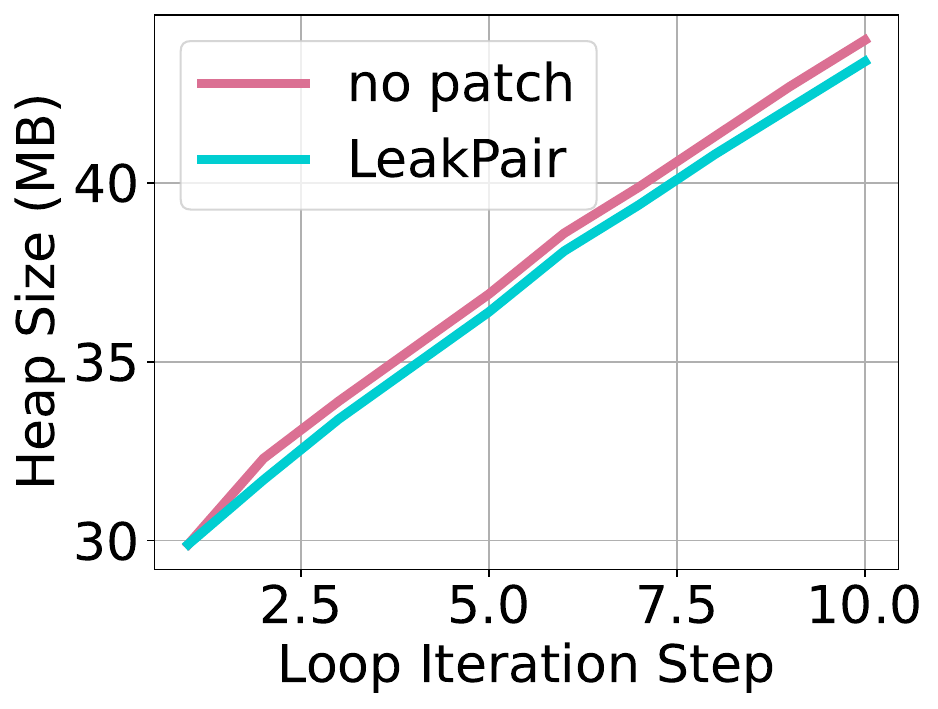}}
    \subfloat[U7]{\includegraphics[width=0.20\linewidth]{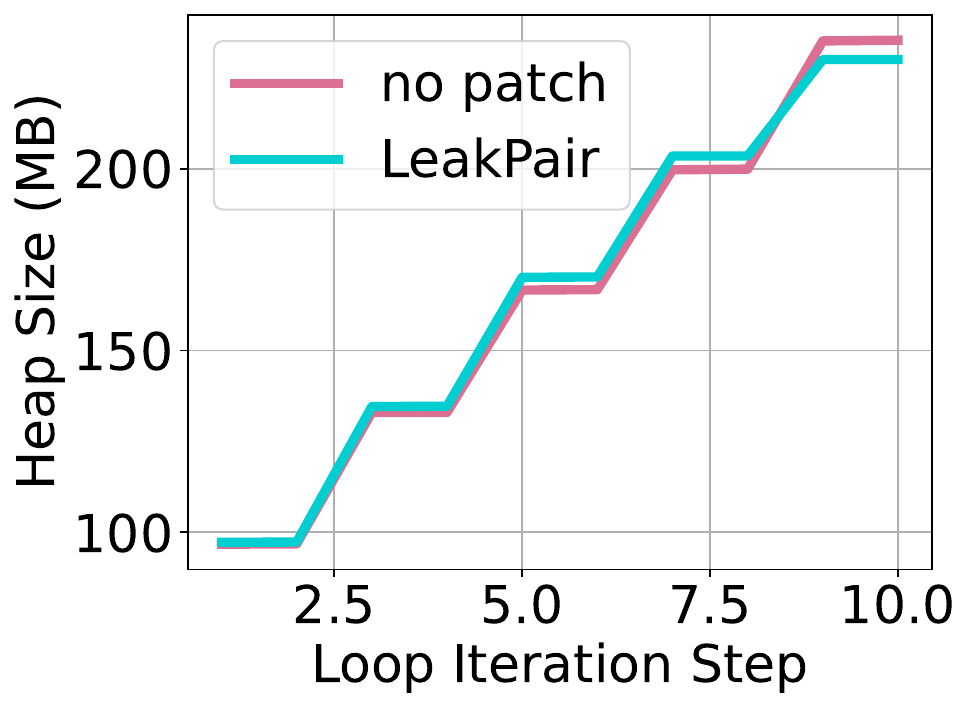}}
    \subfloat[U8]{\includegraphics[width=0.20\linewidth]{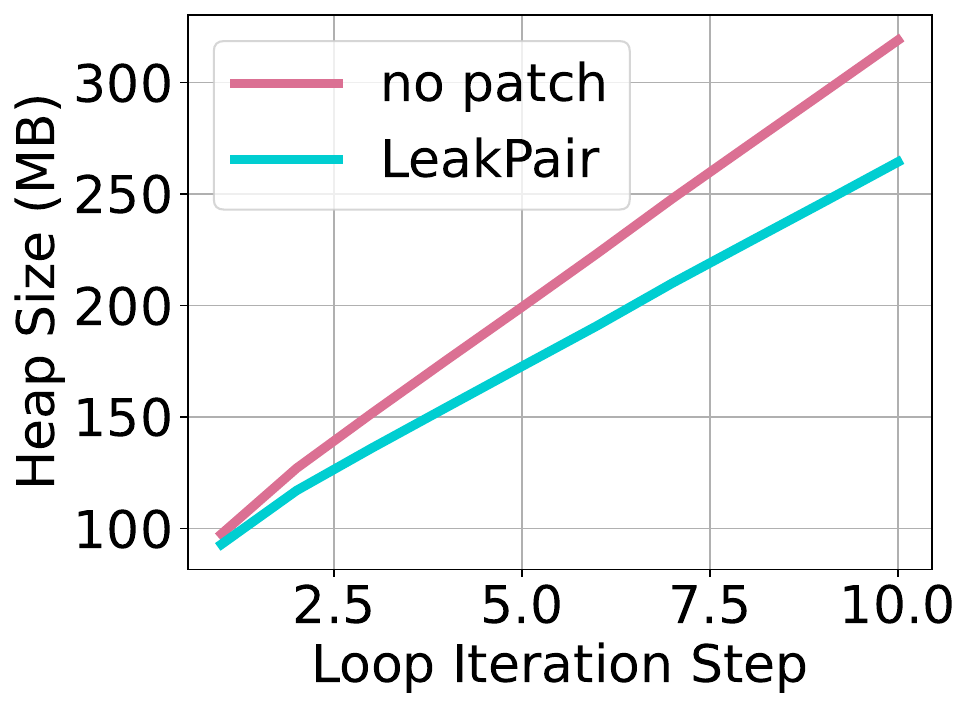}}
    \subfloat[U9]{\includegraphics[width=0.20\linewidth]{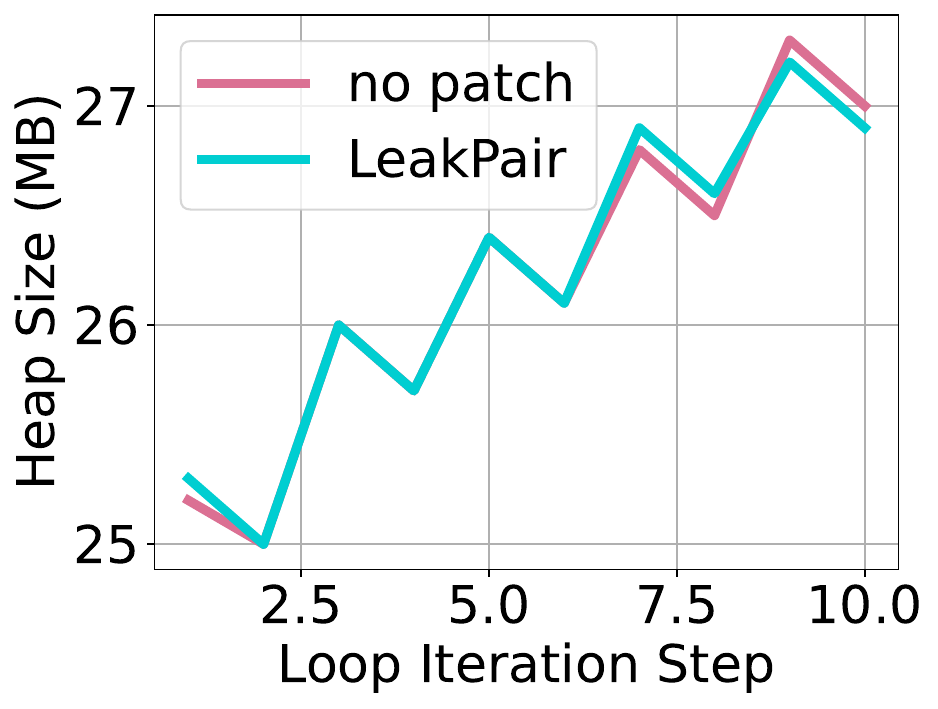}}
    \subfloat[U10]{\includegraphics[width=0.20\linewidth]{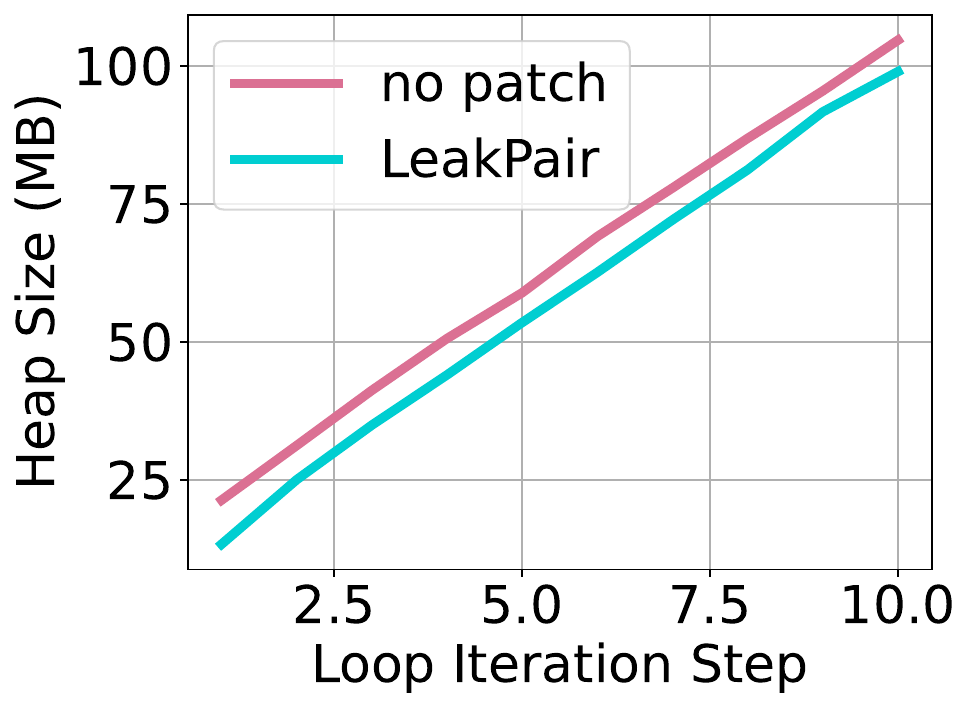}}
    \\
    \subfloat[K1]{\includegraphics[width=0.20\linewidth]{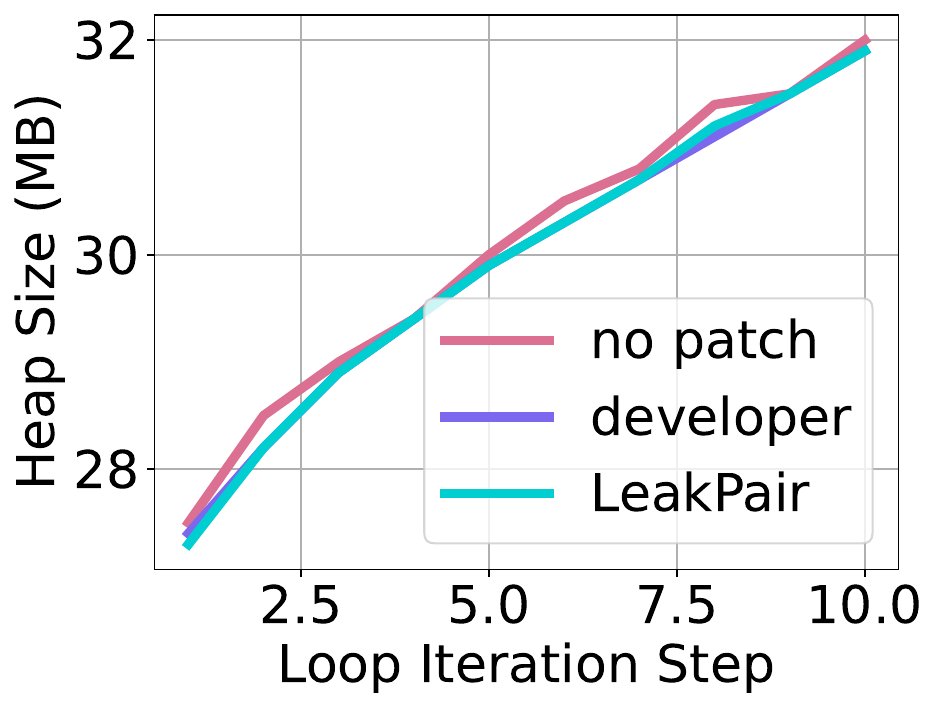}}
    \subfloat[K2]{\includegraphics[width=0.20\linewidth]{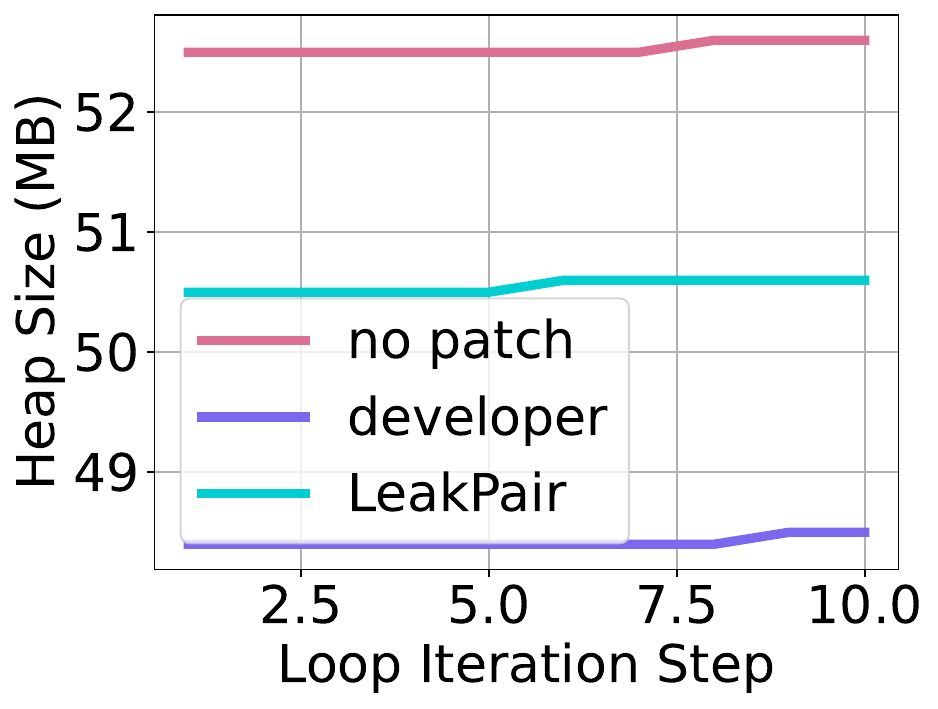}}
    \subfloat[K3]{\includegraphics[width=0.20\linewidth]{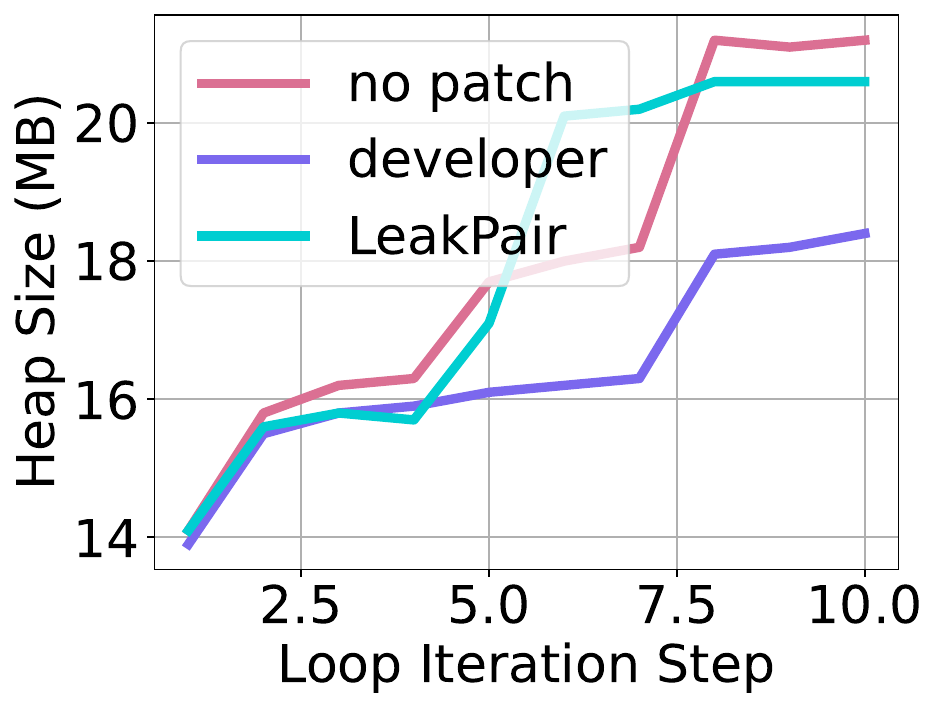}}
    \subfloat[K4]{\includegraphics[width=0.20\linewidth]{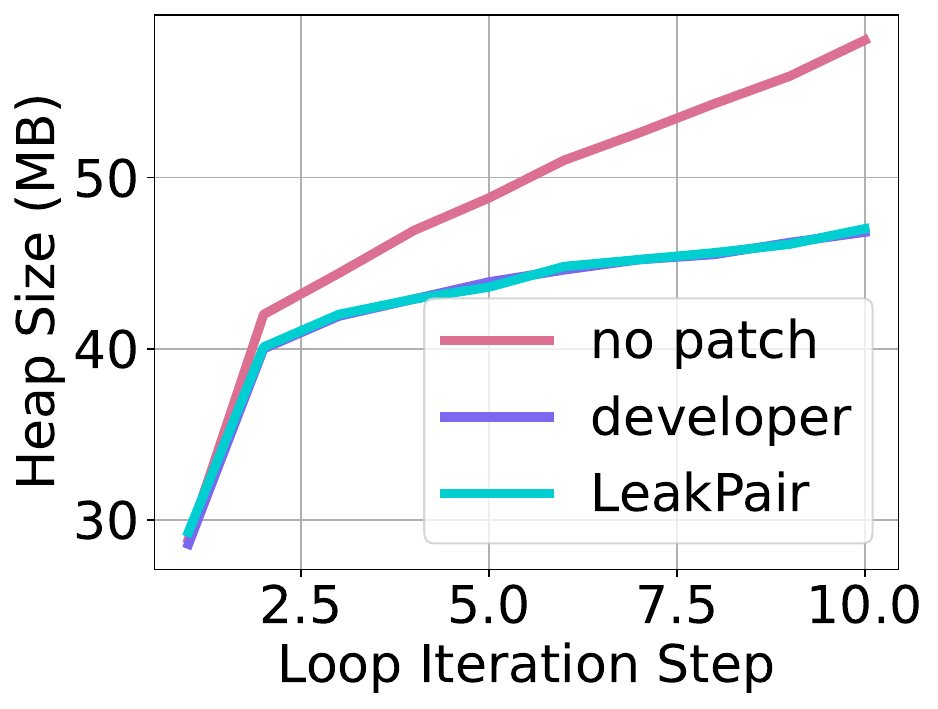}}
    \subfloat[K5]{\includegraphics[width=0.20\linewidth]{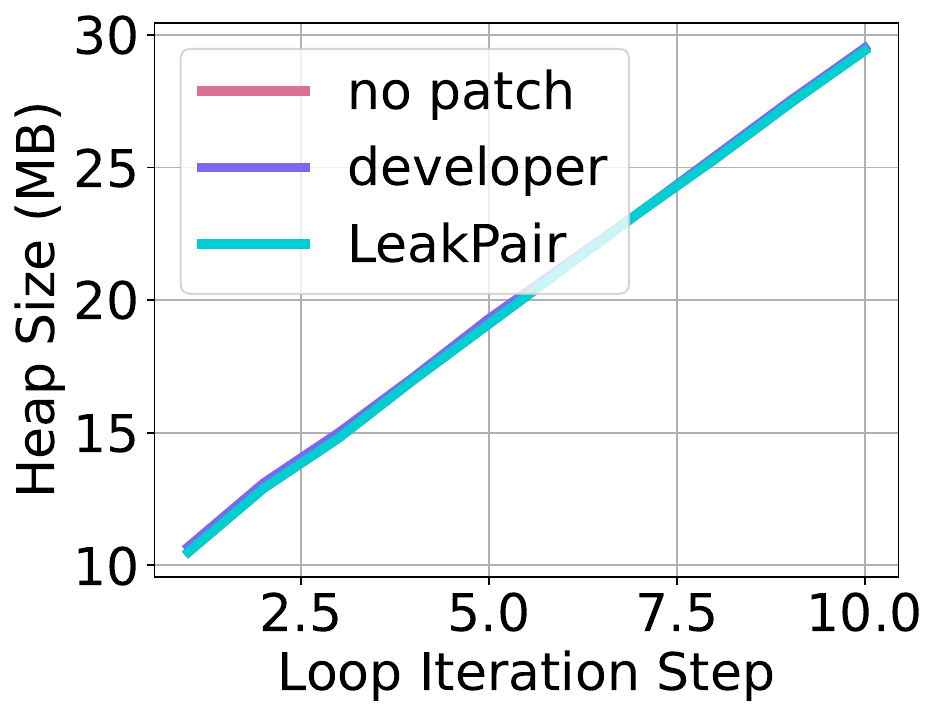}}
    \\
    \subfloat[K6]{\includegraphics[width=0.20\linewidth]{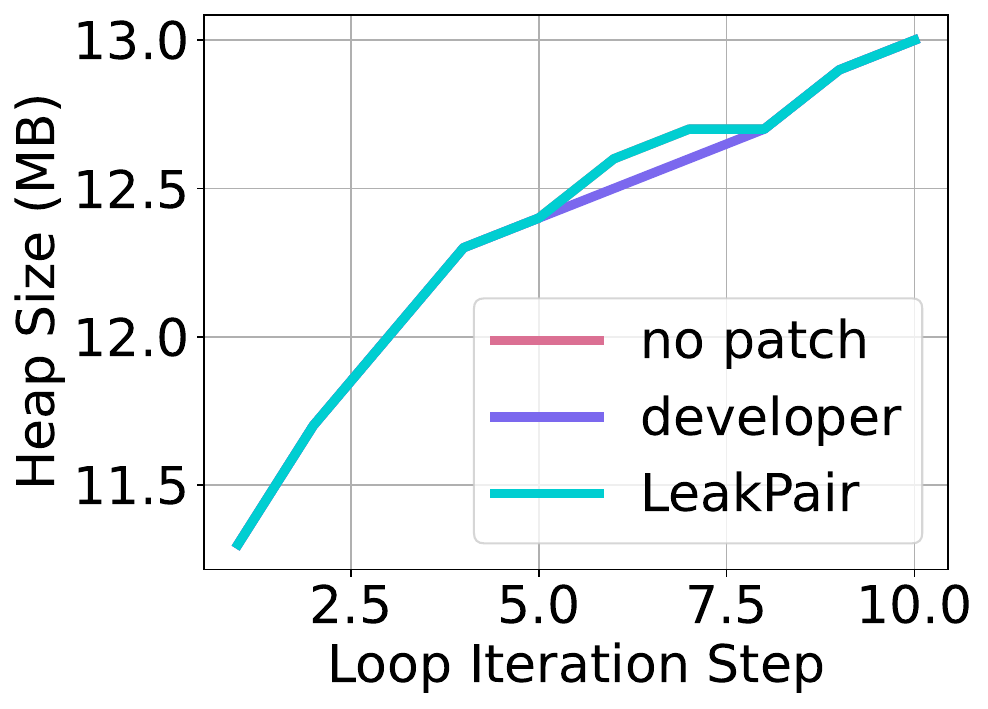}}
    \subfloat[K7]{\includegraphics[width=0.20\linewidth]{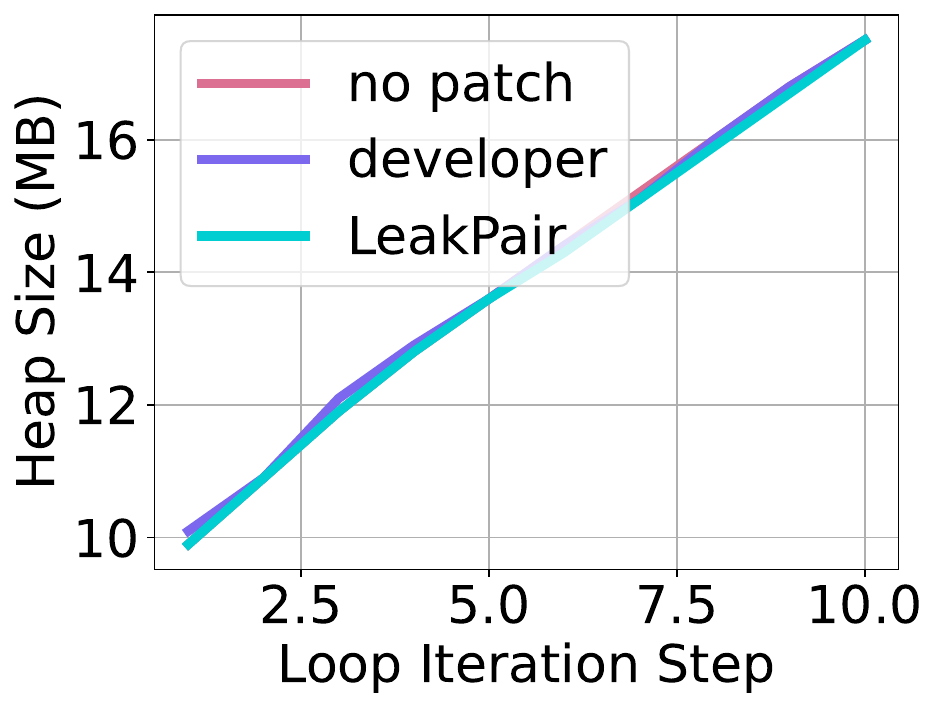}}
    \subfloat[K8]{\includegraphics[width=0.20\linewidth]{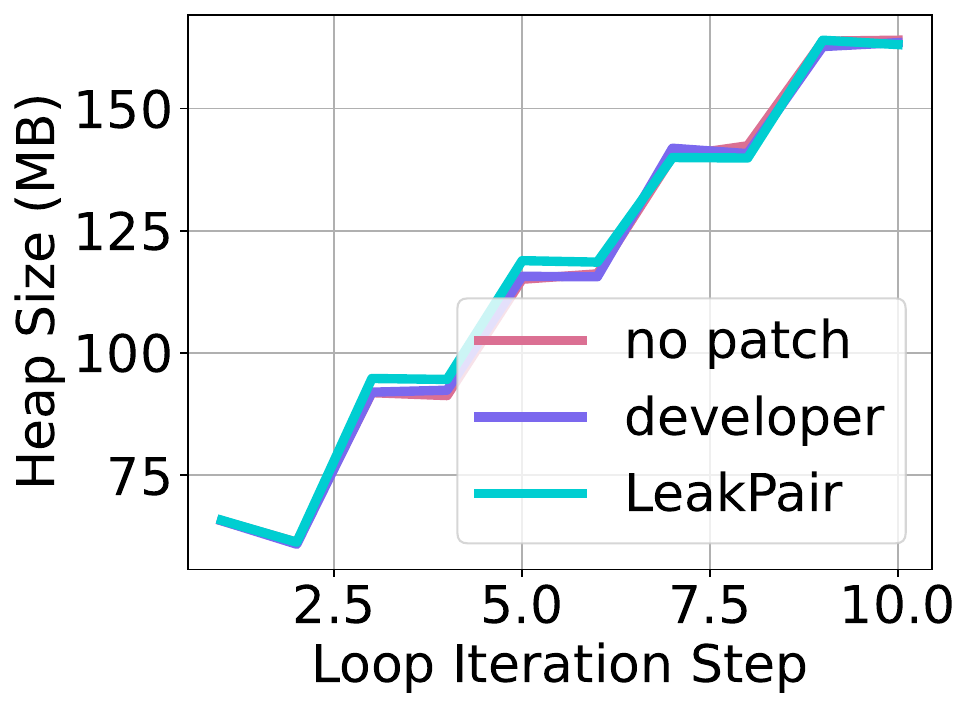}}
    \subfloat[K9]{\includegraphics[width=0.20\linewidth]{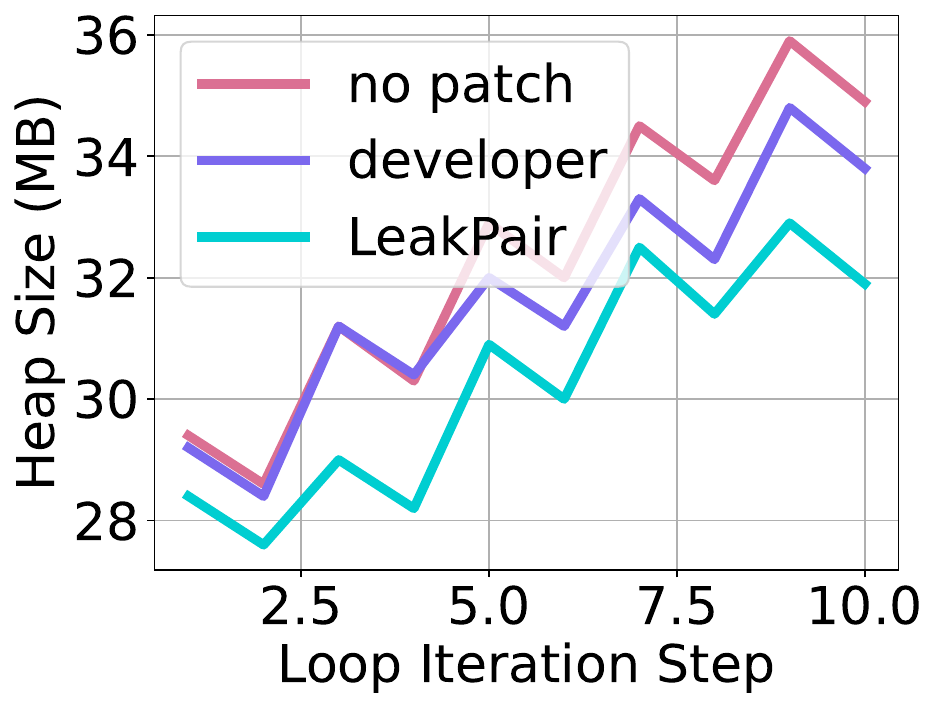}}
    \subfloat[K10]{\includegraphics[width=0.20\linewidth]{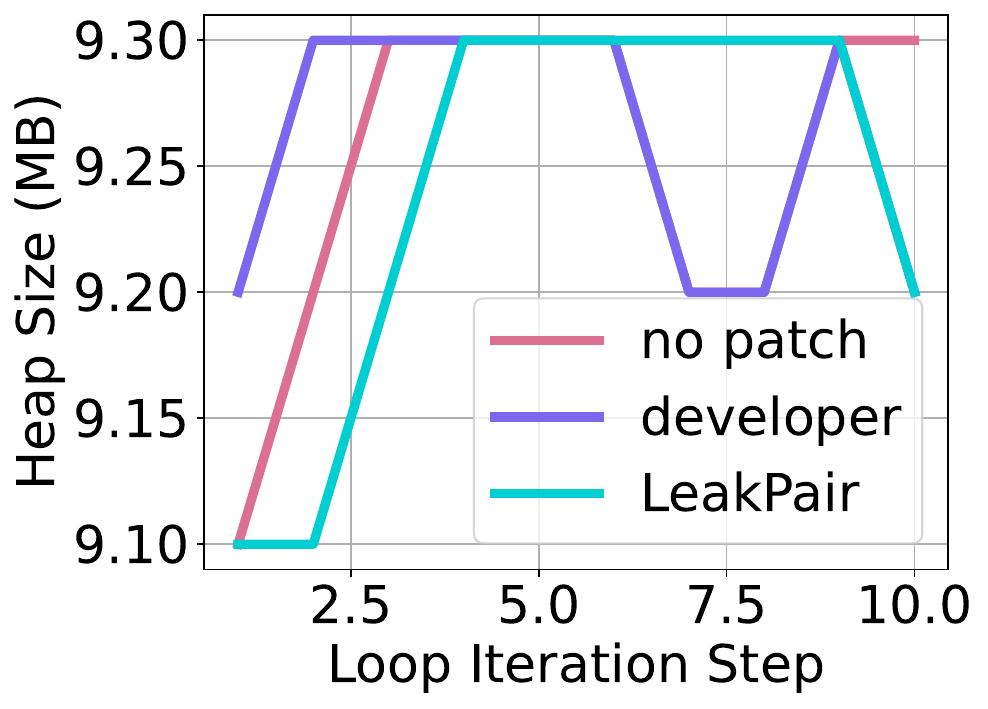}}
    \caption{Heap size over loops after applying \toolname to the subjects listed in Tables~\ref{tab:result:unknown} and~\ref{tab:result:known}.}
    \label{fig:plots}
  }
\end{figure*}

\subsection{RQ1: How effective is \toolname?}

The patches generated by \toolname can reduce memory consumption,
as shown in Tables~\ref{tab:result:known} and~\ref{tab:result:unknown}.
We apply the tool to each subject listed in Tables~\ref{tab:subject:unknown} (projects with
unknown memory leaks) and~\ref{tab:subject:known}
(projects with already known leaks) according to the procedure described in
Section~\ref{sec:setup:rq1}.
In the result tables, the \cc{Leak Patterns} column lists the fix patterns (see Section~\ref{sec:patterns})
successfully applied to each subject.
The \cc{Leaked Objects $\ast$} columns represent the number of clusters in which
objects are potentially leaking the memory space, before and after applying
our tool, and the difference.
The \cc{Heap Size $\ast$} columns show the size of the heap in each subject before and after applying
our tool, and the difference.

As shown in Tables~\ref{tab:result:known} and~\ref{tab:result:unknown},
respectively, \toolname can reduce both memory consumption and potentially
leaking objects.
Their statistical significance of the differences are denoted as
$\ast$:p-value$<$0.05 and $\ast\ast$:p-value$<$0.01.
The reduction is relatively larger for the subjects with unknown leaks.
The higher effectiveness shown in Table~\ref{tab:result:unknown} might indicate that the subjects with unknown leaks paid less attention to memory management while those in Table~\ref{tab:result:known} paid more attention; which is why we
could identify already-known memory leak patches from the subjects.

The patches do not introduce any new leaks, as shown in the plots in Figure~\ref{fig:plots}.
The plots illustrate the changes in the memory heap size in one execution for each subject.
Although there was some fluctuation due to the nature of web applications (e.g.,
it can be affected by the browser status even for the same scenarios),
it turns out that our patches contribute to reducing memory consumption, or at least, they do not add to it. One of the patches (\cc{K9}) reduces the heap size even more than the developer's patch.

\begin{table}[t!]
    \centering
    \footnotesize
    \caption{Results of pull requests reporting the patches generated by \toolname, which fix unknown leaks in subjects listed in Table~\ref{tab:subject:unknown}.}
    \label{tab:pullrequests}
    \resizebox{\linewidth}{!}
    {
    \begin{threeparttable}
        \begin{tabular}{ccc||c|c|c}
            \toprule
            \multicolumn{3}{c|}{\bf Agreed} 
            & \multirow{2}{*}{\bf Disagreed}
            & \multirow{2}{*}{\bf Ignored} & \multirow{2}{*}{\bf Total} \\\cline{1-3}
            Merged & Approved & Improved 
            & & \\
            \hline
             8 & 2 & 1 &  0  & 9  & 20 \\
            \bottomrule
        \end{tabular}
    \end{threeparttable}
    }
\end{table}

The results of our experiments may imply that \toolname
is effective for most SPAs, no matter how it is maintained. It might be helpful to reduce the memory consumption, and it can further prevent potential memory bloats. Furthermore, it does not add any harmful code and does not increase memory consumption in any way.

\find{{\bf Answer to RQ1:} \toolname can generate patches to fix
  memory leaks in SPAs without leak detection, and the patches successfully reduce
  applications' memory consumption. It turns out that they are competitive with the original patches written by human developers.}

\subsection{RQ2: Are the patches by \toolname acceptable?}
\label{sec:acceptability}

To assess the acceptability of patches generated by  \toolname, a live study was carried out on active open-source SPA projects (including SPA websites and libraries used by them), as described in Section~\ref{sec:setup:rq2}.

The study involves creating pull requests (PRs) for patches by \toolname for the subjects in Table~\ref{tab:subject:unknown}, and observing the outcome of pull requests.
We submitted 20 pull requests for 17 of those subjects after clustering similar leaks/patches and confirming a substantial reduction in the count of memory leaks or heap size by the patches, together with the analysis results by Memlab~\cite{memlab}.

Table ~\ref{tab:pullrequests} contains the results of the live study up to the date of the submission. 11 out of 20 PRs (55\%) are approved by the developers, out of which 8 were merged directly. One PR led to the creation of a separate PR by the project developers based on the changes in our PR, which addressed the same leak patterns but used a slightly different approach (in compliance with their specific programming conventions), which was then merged. The leak patterns repaired in 2 of the PRs are approved as anti-patterns by the authors that need to be addressed; however, the PRs for them have not yet been merged. The authors have taken note of our repairs and plan to address the leak patterns themselves in the near future.

One of our PRs inspired the project owner to fix a similar memory leak pattern together with the one in the PR. It is worth noting that no PR has been rejected so far, which further corroborates the non-intrusive nature of \toolname patches. Nine PRs did not get any response from the developers up to the date of submission, however, this might very likely change as the PRs are all fairly recently submitted.




\find{{\bf Answer to RQ2:} The patches generated by \toolname are even acceptable
  to the developers of the target projects. While more than half of the patch
  suggestions are accepted, there are no explicitly rejected patches.}

\begin{table}[!t]
  \centering
	\setlength\tabcolsep{2pt}
	\caption{Test execution applying \toolname to the subjects in Table~\ref{tab:subject:unknown}.}
	\label{tab:test:unknown}
	\resizebox{1.00\linewidth}{!}{%
\begin{threeparttable}
\begin{tabular}{c|c|c|c|c}
\toprule
\multirow{2}{*}{\bf ID} &
Test results before &
Test results after  &
Elapsed time before &
Elapsed time after
\\
 &
 applying \toolname &
 applying \toolname &
 applying \toolname &
 applying \toolname
\\\midrule
U1 & N/A & N/A & N/A & N/A \\
U2 & N/A & N/A & N/A & N/A \\
U3 & 46 passed of 46 & 46 passed of 46 & 8.1 s & 8.4 s \\
U4 & 126 passed of 129 & 126 passed of 129  & 0.3 s & 0.3 s \\
U5 & 6 passed of 14 & 6 passed of 14 & 3.9 s & 1.4 s \\
U6 & 101 passed of 101 & 101 passed of 101 & 55 s & 56.6 s \\
U7 & 66 passed of 66 & 66 passed of 66 & 119.835 s & 120.835 s \\
U8 & 1031 passed of 1038  & 1031 passed of 1038  & 41.623 s & 43.683 s \\
U9 & 43 passed of 43 & 43 passed of 43 & 6.478 s & 6.318 s \\
U10 & 12 passed of 12 & 12 passed of 12 & 0.3 s & 0.3 s
\\\bottomrule
\end{tabular}
{The full list of subjects used for our experiment is available in the replication package~\cite{figshare}.}
\end{threeparttable}
}
\end{table}

\begin{table}[!t]
  \centering
	\setlength\tabcolsep{2pt}
	\caption{Test execution results applying \toolname to the subjects in Table~\ref{tab:subject:known}.}
	\label{tab:test:known}
	\resizebox{1.00\linewidth}{!}{%
\begin{threeparttable}
\begin{tabular}{c|c|c|c|c}
\toprule
\multirow{2}{*}{\bf ID} &
Test results before &
Test results after  &
Elapsed time before &
Elapsed time after
\\
 &
 applying \toolname &
 applying \toolname &
 applying \toolname &
 applying \toolname
\\\midrule
K1 & N/A & N/A & N/A & N/A \\
K2 & N/A & N/A & N/A & N/A \\
K3 & 14 passed of 14 & 14 passed of 14 & 41.2 s & 35 s \\
K4 & N/A & N/A & N/A & N/A \\
K5 & 1610 passed of 1610 & 1610 passed of 1610 & 4 s & 4 s \\
K6 & 64 passed of 64 & 64 passed of 64 & 10.01 s & 10.2 s \\
K7 & 1656 passed of 1750 & 1656 passed of 1750  & 1.5 s & 1.6 s \\
K8 & 275 passed of 277  & 275 passed of 277 & 10.582 s & 8.549 s \\
K9 & N/A & N/A & N/A & N/A \\
K10 & 101 passed of 101 & 101 passed of 101 & 3.685 s & 3.885 s\\\bottomrule
\end{tabular}
{The full list of subjects used for our experiment is available in the replication package~\cite{figshare}.}
\end{threeparttable}
}
\end{table}

\subsection{RQ3: Do the patches break the functionality?}

To show the non-intrusiveness of the patches generated by our tool,
we ran the test cases of each subject according to the procedure
explained in Section~\ref{sec:setup:rq3}. We could not run test suites
for two and four subjects listed in Tables~\ref{tab:subject:unknown} and~\ref{tab:subject:known}, respectively. The tables also report on the
execution time of the test suites as well.

As shown in Tables~\ref{tab:test:unknown} and~\ref{tab:test:known}, the patches generated by \toolname do not introduce any new positive or negative test outcomes.
For subjects with some skipped and failed test cases,
we checked if any new positive or negative test cases had replaced the previous outcomes. As a result, we found no replacement,
which indicates that our patches do not change the behaviors of
the subjects, at least with respect to the test suites provided.
In addition, there were no significant differences with respect to
test execution time as well.

The results of this experiment show that \toolname is unlikely to
break the functionality of SPAs when generating patches to fix
potential memory leaks. This implies that the users of \toolname
may apply the tool without having the functionality changed.
Although running test suites may not guarantee the non-intrusiveness
of patches, our tool is highly likely to generate patches that
preserve the behaviors of the programs.

\find{{\bf Answer to RQ3:} According to the test results, the patches by \toolname are not intrusive. Although test suites cannot guarantee their correctness,
  the patches do not break any functionality, at least from a maintenance perspective.}

\section{Discussion}

\subsection{Comparison against state-of-the-art}
\label{sec:coconut}
To the best of our knowledge, \toolname is the only program repair tool that fixes memory leaks in JavaScript programs.
As of February 2023, the only alternative program repair tool that can deal with Javascript programs is \coconut~\cite{lutellier_coconut_2020}.
\coconut is a recent learning-based program repair tool. Since its training data contains Javascript programs (3,217,093 programs obtained from 10,163 open-source projects), it can be used to fix Javascript programs.
As a general-purpose APR tool, \coconut can also be used to fix memory leaks.

To assess the performance of \coconut in fixing memory leaks, we apply \coconut to all pairs of buggy and fixed versions from which we mined our fix patterns.
\coconut, like most program repair tools, requires buggy lines, which we provide with ground truth patches.
In case the ground-truth fix modifies multiple lines, we apply \coconut to each of those lines.
Then, at each buggy line, we compare the ground-truth fix and the 1,000 \coconut-generated fixes (\coconut uses beam search with a beam width 1,000).
In \uline{none} of the buggy lines, \coconut generates a ground-truth fix.

\begin{figure}[h!]
\begin{center}
\includegraphics[width=0.5\linewidth]{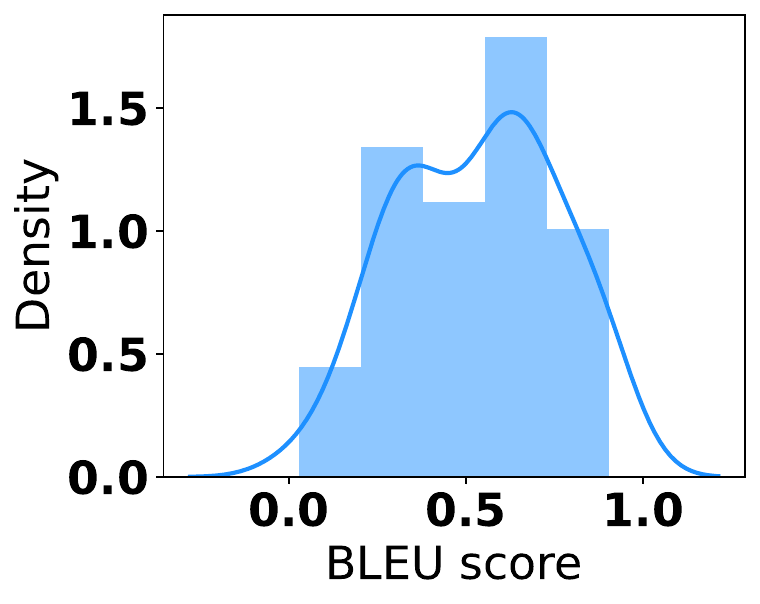}
\end{center}
\caption{Distribution of the BLEU scores of \coconut-generated patches.}
\label{fig:coconut-bleu}
\end{figure}

To assess \coconut at a finer-granularity level (token level), we compute BLEU scores at each line.
Figure~\ref{fig:coconut-bleu} shows the distribution of the obtained BLEU scores. At each line, we consider only the maximum score out of 1,000.

Our \coconut-experiment results suggest that (1) the learning-based approach such as \coconut shows the potential to generate a correct fix (the highest BLEU score is 0.90), but (2) the performance is not as good as being able to generate a correct fix (not a single version is correctly fixed).
While general-purpose program repair is attractive, it is not a panacea.
For bugs that can be fixed with patterns, such as memory leaks, our pattern-based approach works better and more reliably.

\subsection{Threats to Validity}

\textbf{Threats to external validity}
may lie in the target subjects that this study uses as they are open-source projects;
thus, the results may not be representative of projects, such as those using closed-source techniques. In addition, our study focuses only on JavaScript subjects, while
there are other languages implementing SPAs.
This threat might be mitigated since our target SPA frameworks (i.e., React and Angular)
are popular and representative in the web development community.

\noindent
\textbf{Threats to internal validity}
may include fix patterns manually extracted by the authors.
To address this threat, each fix pattern is supported by
real patches that fix memory leaks in SPAs implemented by React and Angular.

\noindent
\textbf{Threats to construct validity}
may relate to the test cases used in the evaluation.
To show the non-intrusiveness of the patches generated by \toolname,
our experiment runs test cases given by each subject.
Although test suites may not prove the correctness of the behavior
in the applications, it might be enough to preserve major functionalities
in the applications from the maintenance perspective.

\section{Related Work}
\label{sec:related}


\subsection{Pattern-based Program Repair}
Program repair with fix patterns (or fix templates) was first introduced by Kim
et al.~\cite{6606626}, where the authors manually inspected 60,000+ human-
written patches. Based on the inspection, common patterns were derived that
were then implemented as automated fix templates in PAR (Patch-Based Automated
Program Repair). The tool was evaluated by applying it to 119 real-world bugs
and comparing the number of patches generated by PAR that were approved by 253
human subjects with those generated by GenProg~\cite{weimer_automatically_2009}.
Patches generated by PAR were shown to have a higher acceptance ratio.

Pattern-based program repair has been improved and leveraged with
many other ideas. There are studies that extract patterns for different
targets such as  JavaScript faults~\cite{Vejovis} and performance
bugs~\cite{nistorcaramel2015}. Researchers leveraged diverse
sources of fix patterns, for example, Q\&A posts~\cite{8330202},
similar snippets~\cite{3213871}, fault localization results~\cite{fixminer},
and static analysis warnings~\cite{avatar}. In addition, TBar~\cite{tbar}
incorporated common fix patterns from other existing studies and showed that
fix patterns are effective when fixing bugs.

Fix patterns are also utilized to generate non-intrusive patches as well.
The authors of Caramel~\cite{nistorcaramel2015} examined patches submitted
to fix performance bugs in open-source projects written in C and Java languages.
The technique identifies potential performance issues in a program.
and generates patches that do not change the functionality of the program,
i.e., non-intrusive fixes.

\subsection{Memory Leak Debugging}

There have been some techniques proposed to address memory leaks in Javascript projects.
Qian et al.~\cite{Lightweight} proposed a technique that reports the suspected
leaking objects by collecting the application heap snapshots and using a
lightweight statistical algorithm that combines several heuristics.
BLeak~\cite{bleak} is based on the notion that web app users often return to
the same visual state after performing some actions. The rationale is that
visiting the same state should consume the same amount of memory; therefore, if
there is sustained growth in memory consumption between the loops to the same
state, it is a valid indicator of memory leakage.

Another common line of research involves dynamic approaches.
One dynamic approach for non-garbage-collected languages, presented by Azhari
et al.~\cite{9671473} performs memory leak detection by memory block growth analysis.
Another dynamic leak detector, DEF\_LEAK~\cite{defleak}, employs symbolic
execution to detect memory leaks across all paths of execution.
LeakSpot~\cite{leakspot} addresses memory leaks in JavaScript by leveraging a
heap snapshot model. MemInsight~\cite{meminsight} instruments the JavaScript
code and provides a detailed analysis of the applications’ memory dynamics.
Memory Validator~\cite{softwareverify} is a popular memory leak and memory
errors detection tool for C, C++, C\#, Visual Basic and Fortran.

Recently, memory leak detection techniques have leveraged neural networks.
MVD~\cite{mvd} makes use of a novel kind of graph neural network called flow-
sensitive graph neural network (FS-GNN). FS-GNN helps capture critical
contextual data of the code by embedding both statements and flow information
in order to learn program semantics. This model can be trained to learn
vulnerability patterns from the source code as well as detect statements that
are suspected to be vulnerable. It does so by incorporating semantic
information such as call relations and return values from Call graphs into the
basic Program Dependence Graph (PDG).

\section{Conclusion}

In this work, we have introduced a novel technique \toolname to fix memory leaks in single page web applications.
Despite the prevalence of single-page web applications and their memory leaks, there has been no research effort to fix those bugs automatically. We have shown that by using only a handful of fix patterns mined from the existing patches, diverse SPAs of 37 open-source projects can be successfully fixed. Furthermore, the patches generated by \toolname are high-quality (the majority of the pull requests \toolname made were accepted by the original developers) and safe to accept (the fix patterns we use are non-intrusive).


This work also aims at fixing a specific type of bug, i.e., memory leaks in single-page applications.
The proposed technique is simple as compared to recent approaches.
However, simplicity does not necessarily imply ineffectiveness.
On the contrary, \toolname is very effective, as was shown.
We view this as the strength of our approach.
For certain types of bugs, simple pattern-based approaches, like ours, do a good job without using heavy-weight deep learning or implementing complex static analysis and proving the correctness of the analysis.

\section{Data Availability}
We make the replication package publicly available, which includes all the code and datasets to reproduce our experiments at \url{https://figshare.com/s/5991a6f89800906176a2}~\cite{figshare}.

\section*{Acknowledgments}
This work was supported by the National Research Foundation of Korea (NRF) grant funded by the Korea government (MSIT) (No. 2021R1A5A1021944 and 2021R1I1A3048013) and the Institute for Information \& Communications Technology Planning \& Evaluation (IITP) grant funded by the Korea government (MSIT) (No.2021-0-01001). Additionally, the research was supported by Kyungpook National University Research Fund, 2020.


\balance
\bibliographystyle{IEEEtran}
\bibliography{bib/main}


\end{document}